\documentclass[twocolumn,pre,showpacs]{revtex4}
\usepackage{amssymb}
\usepackage{amsfonts}
\usepackage{amsmath}
\usepackage{graphicx}

\begin{document}

\title{Central limit theorem for a class of globally correlated random
variables}
\author{Adri\'{a}n A. Budini}
\affiliation{Consejo Nacional de Investigaciones Cient\'{\i}ficas y T\'{e}cnicas
(CONICET), Centro At\'{o}mico Bariloche, Avenida E. Bustillo Km 9.5, (8400)
Bariloche, Argentina, and Universidad Tecnol\'{o}gica Nacional (UTN-FRBA),
Fanny Newbery 111, (8400) Bariloche, Argentina}
\date{\today }

\begin{abstract}
The standard central limit theorem with a Gaussian attractor for the sum of
independent random variables may lose its validity in presence of strong
correlations between the added random contributions. Here, we study this
problem for similar interchangeable globally correlated random variables.
Under these conditions, a hierarchical set of equations is derived for the
conditional transition probabilities. This result allows us to define
different classes of memory mechanisms that depend on a symmetric way on all
involved variables. Depending on the correlation mechanisms and single
statistics, the corresponding sums are characterized by distinct statistical
probability densities. For a class of urn models it is also possible to
characterize their domain of attraction which, as in the standard case, is
parametrized by the probability density of each random variable. Symmetric
and asymmetric $q$-Gaussian attractors $(q<1)$ are a particular case of
these models.
\end{abstract}

\maketitle



\section{Introduction}

The standard central limit theorem (CLT) is a cornerstone of probability
theory \cite{kolmogorov,feller,vanKampen,gardiner}. It establishes that a
sum of independent (identical) random variables, under a suitable rescaling,
converges to a Gaussian distribution. It plays a fundamental role in the
formulation of statistical thermodynamics and also provide a rigorous basis
for assuming Gaussian statistics for describing fluctuations in equilibrium
and nonequilibrium systems.

There exist a few remarkable examples where the standard CLT was
generalized. The Gaussian attractor arises when considering independent
random variables with a finite second moment. As is well known, when this
condition is raised up the attractor becomes a Levy distribution \cite{levy}%
. On the other hand, Gumbel distribution arises from the study of extreme
value statistics and describes the fluctuations of the largest value in a
large set of identically distributed independent random variables \cite%
{gumbel}. Interestingly, this problem can in general be related with the
statistics of random sums of correlated random variables \cite{bertin}.
Departure from Gaussian statistics was also analyzed for global correlations
where the characteristic function of the total sum is defined by a
non-multiplicative Fourier structure \cite{baldowin}.

Recently it was argued that the presence of global correlations in
stationary equilibrium and nonequilibrium systems is a situation where
nonextensive statistical mechanics may applies \cite{TsallisBook,rapisarda,
tirnakli,ruseckas}. Consistently, many theoretical effort was devoted to
finding global memory mechanisms that lead to attractors defined by $q$%
-Gaussian probability densities \cite%
{hilhorst,Qhanel,QCentralLimit,rodri,Qurnas,plastino}. These statistical
objects also arise from maximizing Tsallis entropy \cite{TsallisBook}, from
superstatistical models \cite{superstatistics}, as well as from specific
transformations of Gamma distributed random variables \cite{budini}.

Global correlations are a mechanism that may lead to departures from
Gaussian statistics. Nevertheless, establishing a generalization of the CLT
on the basis of\ only this feature is a formidable task. In fact, to our
knowledge, there not exist general rigorous mathematical criteria for
splitting correlations in weak ones (leading to Gaussian statistics) and
stronger ones (departure from normal distribution). Therefore, as in the
previous literature \cite%
{bertin,baldowin,TsallisBook,rapisarda,tirnakli,ruseckas,Qhanel,QCentralLimit,rodri,Qurnas,plastino,hilhorst}%
, one is naturally forced to study particular cases. Of special interest is
to find generalizations that rely on simple correlation mechanisms or
symmetries, which in turn also allow defining or studying its domain of
attraction. In general, this last issue is hard to solve.

In this paper we analyze the departure from the standard CLT for a specific
class of global correlations. Similar interchangeable random variables \cite%
{finetti,jaynes,hewit,heath,kingman} are considered. This property or
symmetry, originally introduced by de Finetti \cite{finetti} in probability
theory, is defined by random variables whose joint probability density is
invariant under arbitrary permutations of its arguments.

The main goal is twofold. First, we give a general characterization of
possible correlations mechanisms consistent with interchangeability. This
objective is achieved by characterizing the correlations not through the
joint probability densities but through the transition probabilities. These
functions say us how the probability density of a given variable depends on
the previous values assumed by the rest of the random variables. We
demonstrate that these objects obey a set of hierarchical equations that
resemble a Chapman-Kolmogorov equation for Markovian chains \cite%
{feller,vanKampen,gardiner}. From this result we construct different
correlation models which allow us to achieve the second main goal, that is,
the characterization of the departure from Gaussian statistics as well as to
study their domain of attraction. For a class of urn models \cite%
{norman,pitman,queen,restaurante}, we demonstrate that their domain of
attraction is as wide as in the standard case. Asymmetric and symmetric $q$%
-Gaussian attractors \cite{budini}\ with $q<1$ arise as a particular case of
these urn models.

The paper is outlined as follows. In Sec. II, based on the
interchangeability property of the joint probabilities, we derive a
hierarchical set of equations for the transition probabilities. Sec. III is
devoted to finding different solutions to the previous equations, which are
based on a generalization of P\'{o}lya urn scheme \cite%
{norman,pitman,queen,restaurante}. In Sec. IV, departure from Gaussian
statistics and their basin of attraction are analyzed. In Sec. V we provide
the Conclusions. In the Appendixes we show some calculus details and study
other correlation models (additive memory, de Finetti representation,
Blackwell-MacQueen urn scheme).

\section{Hierarchy of transition probabilities for similar interchangeable
random variables}

A set of $n$ random variables $X_{1},X_{2},\cdots X_{n},$ can be
characterized by the $n$-joint probability distribution $P_{n}(x_{1,}x_{2},%
\cdots x_{n}),$ which defines the probability that each variable falls in an
infinitesimal range $dx_{i}$ around $x_{i}.$

Similar interchangeable variables are defined by the following two
symmetries. Similarity (or scale invariance \cite{TsallisBook}) means that
for any $n$ it is fulfilled the relation%
\begin{equation}
P_{n-1}(x_{1},x_{2},\cdots x_{n-1})=\int dx_{n}P_{n}(x_{1},x_{2},\cdots
x_{n}).
\end{equation}%
Therefore, the joint probability density of the subset of $(n-1)$ random
variables coincides with the marginal distribution corresponding to $n$
variables. On the other hand, interchangeability is defined by the
invariance of the joint probability density under arbitrary permutations of
its arguments,%
\begin{equation}
P_{n}(\cdots ,x_{k},\cdots ,x_{l},\cdots )=P_{n}(\cdots ,x_{l},\cdots
,x_{k},\cdots ),  \label{symmetry}
\end{equation}%
that is, for any $k$ and $l$ in the interval $(1,2,\cdots n),$ the joint
probability density does not change under the (arbitrary) interchange $%
x_{k}\leftrightarrow x_{l}.$ These relations are assumed valid for all
values of $n.$ Notice that in particular the previous two conditions imply
that all random variables $\{X_{i}\}_{i=1}^{n}$ are characterized by the
same distribution, $P_{1}(x).$

The joint probability density $P_{n}(x_{1,}x_{2},\cdots x_{n})$ completely
characterizes the random variables $\{X_{i}\}_{i=1}^{n}.$ Nevertheless, an
extra aspect is lighted by introducing a conditional probability density
defined by the relation%
\begin{equation}
P_{n}(x_{1},\cdots x_{n})=P_{n-1}(x_{1},\cdots x_{n-1})T_{n-1}(x_{1},\cdots
x_{n-1}|x_{n}).  \label{TransitionNBayes}
\end{equation}%
Hence, the function $T_{n-1}(x_{1},\cdots x_{n-1}|x_{n})$ gives the
probability density of the variable $X_{n}$ \textit{given} that the previous
ones $\{X_{i}\}_{i=1}^{n-1}$ assumed the values $x_{1},\cdots x_{n-1}.$ By
definition, it satisfies the normalization condition $\int
dx_{n}T_{n-1}(x_{1},\cdots x_{n-1}|x_{n})=1.$

From Eq. (\ref{TransitionNBayes}), iteratively it follows%
\begin{eqnarray}
P_{n}(x_{1},\cdots x_{n})
&=&P_{1}(x_{1})T_{1}(x_{1}|x_{2})T_{2}(x_{1},x_{2}|x_{3})\cdots  \notag \\
&&\cdots \times T_{n-1}(x_{1},\cdots x_{n-1}|x_{n}).  \label{Conjunta}
\end{eqnarray}%
Therefore, the set of functions $T_{k}(x_{1},\cdots x_{k}|x_{k+1}),$ with $%
k=1,\cdots n-1$ provide the same information than the $n$-joint probability
density. Furthermore, from Eq. (\ref{Conjunta}) one can easily read how the
correlations between the random variables are build up. In fact, having an
explicit expression for the transition probabilities it is possible to
numerically generate the corresponding realizations of the correlated
variables $\{X_{i}\}_{i=1}^{n}.$

The main problem that we solve in this section is to determine which set of
transition probabilities are consistent with the similarity and
interchangeability properties. \textit{Given} an arbitrary distribution $%
P_{1}(x_{1}),$ the symmetry does not impose any condition. For $n=2,$
interchangeability implies $P_{2}(x_{1,}x_{2})=P_{2}(x_{2,}x_{1}),$ or
equivalently $P_{1}(x_{1})T_{1}(x_{1}|x_{2})=P_{1}(x_{2})T_{1}(x_{2}|x_{1}).$
After integration, and by using the similarity property, it follows the
condition%
\begin{equation}
\int dx_{1}P_{1}(x_{1})T_{1}(x_{1}|x_{2})=P_{1}(x_{2}).
\label{CondicionalOne}
\end{equation}%
By using a similar procedure, $T_{2}(x_{1},x_{2}|x_{3})$ must to fulfill%
\begin{equation}
T_{1}(x_{1}|x_{3})=\int dx_{2}T_{1}(x_{1}|x_{2})T_{2}(x_{1},x_{2}|x_{3}).
\end{equation}%
For higher conditional probabilities densities (see Appendix \ref{jerarquia}%
), the following relations%
\begin{eqnarray}
T_{n-1}(x_{1}\cdots x_{n-1}|x_{n+1}) &=&\int dx_{n}T_{n-1}(x_{1},\cdots
x_{n-1}|x_{n})  \notag \\
&&\times T_{n}(x_{1},\cdots x_{n}|x_{n+1}),  \label{RecursivaCondicional}
\end{eqnarray}%
must to be fulfilled for all values of $n.$ Furthermore, the function $%
T_{n}(x_{1}\cdots x_{n}|x_{n+1})$ must to be \textit{symmetric} in the
conditional arguments $x_{1}\cdots x_{n},$ that is, it is invariant under
arbitrary permutations of its arguments. The hierarchical set of equations
defined by (\ref{RecursivaCondicional}) is the main result presented in this
section.

If $T_{n-1}(x_{1},\cdots x_{n-1}|x_{n+1}),$ for all values of $n,$ does not
depends on the previous values $x_{1},\cdots x_{n-1},$ it follows $%
T_{n}(x_{1},\cdots x_{n}|x_{n+1})=P_{1}(x_{n+1}),$ that is, we recover the
case of \textit{independent identical random variables}. Notice that
interchangeability implies that $T_{n-1}(x_{1}\cdots x_{n-1}|x_{n})$ depends
symmetrically on the previous arguments $x_{1}\cdots x_{n-1}.$ Therefore,
transition probabilities that only depend on one previous value, with a
dependence that is independent on the number of previous events, $%
T_{n-1}(x_{1}\cdots x_{n-1}|x_{n})=T_{n-1}(x_{n-1}|x_{n})=T(x_{n-1}|x_{n}),$
are \textit{not consistent} with interchangeability. This case corresponds
to \textit{stationary Markov chains}. In fact, the unique transition
probability\ $T(x|y)$ should to satisfy [Eq. (\ref{CondicionalOne})]%
\begin{equation}
\int dxP_{1}(x)T(x|y)=P_{1}(y),  \label{M1}
\end{equation}%
while by imposing the previous conditions on Eq. (\ref{RecursivaCondicional}%
), it follows%
\begin{equation}
T(x|y)=\int dx^{\prime }T(x|x^{\prime })T(x^{\prime }|y).  \label{M2}
\end{equation}%
The stationary property is given by Eq. (\ref{M1}), while the Markov
property is defined by the \textit{Chapman-Kolmogorov} relation Eq. (\ref{M2}%
). In fact, the next value (future) depends on the previous value (present
state), but not on the manner in which the present state has emerged from
previous ones (the past).

In the following section we search solutions of Eq. (\ref%
{RecursivaCondicional})\ where the transition probabilities are based on a P%
\'{o}lya urn scheme. In the Appendixes we studied other solutions that also
depend in the same manner on all previous values taken by the random
variables, that is, global correlations. For example an additive memory
assumption $T_{n}(x_{1}\cdots x_{n}|x_{n+1})=\mathcal{T}_{n}(x_{1}+x_{2}%
\cdots +x_{n}|x_{n+1})$ (Appendix \ref{additive}) leads to consistent
solutions for Gaussian and classical spin variables. A generalized de
Finetti representation is analyzed in Appendix \ref{deFinetti}.

\section{Urn schemes}

Urn models are examples of random variables defined through their transition
probabilities \cite{feller,norman}. P\'{o}lya urn scheme generate
interchangeable random variables \cite{pitman,queen,restaurante}. Below we
review this scheme, which gives us the basis for constructing a
generalization consistent with interchangeability.

\subsection{P\'{o}lya Urn scheme}

The standard P\'{o}lya urn scheme can be seen as a particular case of the
Blackwell-MacQueen urn scheme \cite{pitman,queen}, which in turn is related
to the \textquotedblleft Chinese restaurant process\textquotedblright\ \cite%
{pitman,restaurante}. In the present context, it is defined by an arbitrary
distribution $P_{1}(x),$\ while the transition probabilities are%
\begin{equation}
T_{n}(x_{1},\cdots x_{n}|x)=\frac{\lambda P_{1}(x)+\sum_{i=1}^{n}\delta
(x-x_{i})}{n+\lambda }.  \label{TransitionQueen}
\end{equation}%
Here, $\lambda $ is dimensionless positive parameter, while $\delta (x)$ is
the delta Dirac function. When $\lambda \rightarrow \infty ,$ \textit{%
identical independent random variables} are recovered, while the limit $%
\lambda \rightarrow 0$ leads to a \textit{fully correlated case}, that is,
after the first random value the next ones assume the same value.

After a simple algebra it is possible to proof that the set of functions
defined by Eq. (\ref{TransitionQueen}) satisfy Eq. (\ref{CondicionalOne}),
as well as the hierarchical set of conditions corresponding to
interchangeability, Eq. (\ref{RecursivaCondicional}). In Appendix \ref%
{MacQueen}, we analyze the departure from the standard CLT for this model.%
\begin{figure}[tbp]
\includegraphics[bb=60 836 687 1103,angle=0,width=8.6cm]{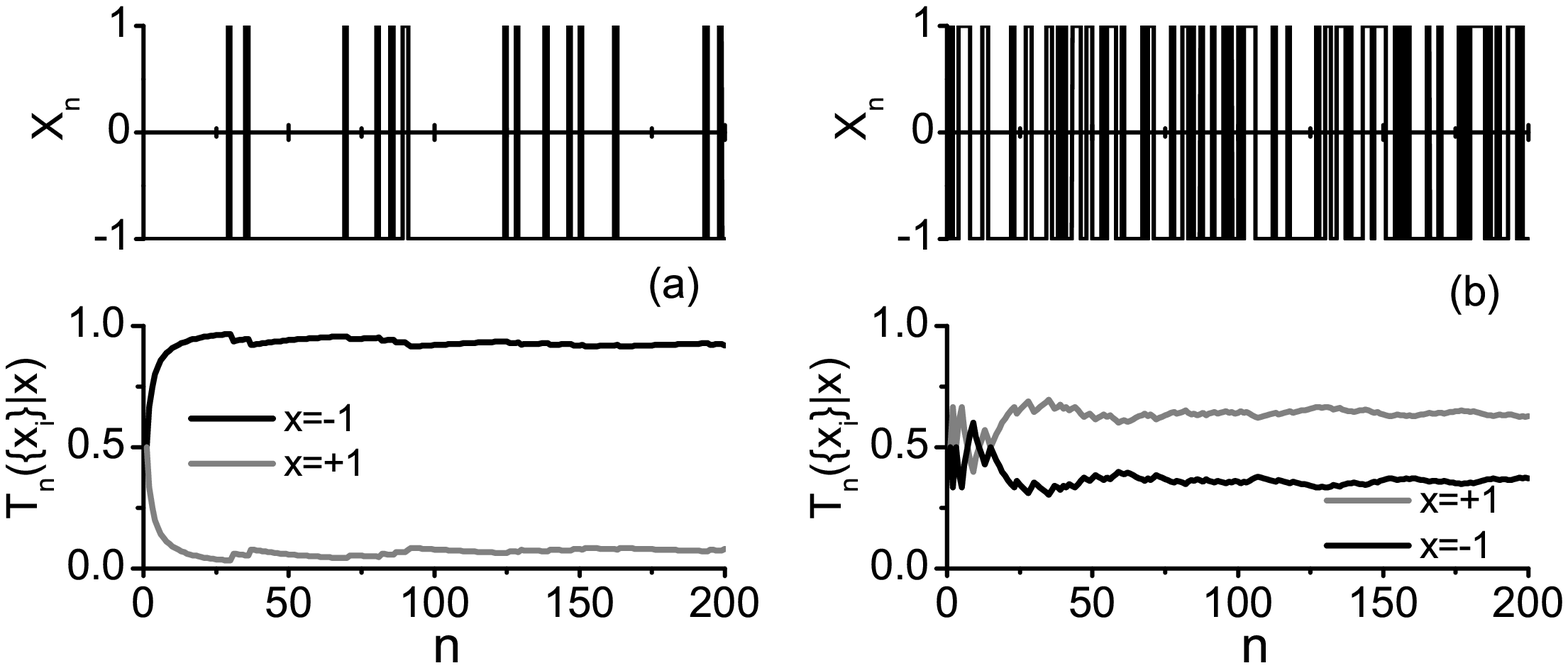}
\caption{Two realizations [(a) and (b)] for a set of classical spin
variables $\{x_{\protect\mu }\}=\{+1,-1\}$ obtained from the Eq. (\protect
\ref{TransitionDirichlet}) $(M=2).$ The lower panels correspond to the
transition probabilities. The parameters are $q_{+}=q_{-}=1/2$ and $\protect%
\lambda =2.$}
\end{figure}

The P\'{o}lya urn scheme corresponds to the particular case in which the
random variables $\{X_{i}\}$ are discrete. Hence, we write%
\begin{equation}
P_{1}(x)=\sum\nolimits_{\mu =1}^{M}q_{\mu }\delta (x-x_{\mu }),
\label{Discreta}
\end{equation}%
where $\{x_{\mu }\}_{\mu =1}^{M}$ is the set of $M$ possible values and $%
\{q_{\mu }\}_{\mu =1}^{M}$ are the corresponding weights (probabilities),
with $\sum\nolimits_{\mu =1}^{M}q_{\mu }=1.$ In this case, the transition
probabilities Eq. (\ref{TransitionQueen}) can be written in terms on the
number of times $n_{\mu }$ that each value $x_{\mu }$ was assumed previously,%
\begin{equation}
T_{n}(\{x_{i}\}|x)=\sum\nolimits_{\mu =1}^{M}\frac{\lambda q_{\mu }+n_{\mu }%
}{n+\lambda }\delta (x-x_{\mu }),  \label{TransitionDirichlet}
\end{equation}%
where $\{x_{i}\}\equiv x_{1},\cdots x_{n}.$ Notice that the set of numbers $%
\{n_{\mu }\}_{\mu =1}^{M}$ that the random values $\{X_{i}\}_{i=1}^{n}$
assumed the values $\{x_{\mu }\}_{\mu =1}^{M}$ satisfy the relation $%
n=\sum\nolimits_{\mu =1}^{M}n_{\mu }.$

The correlation mechanism associated to Eq. (\ref{TransitionDirichlet}) can
be read in the following way. With probability $\lambda /(n+\lambda )$ the
random variable $X_{n+1}$ is draw randomly in agreement with the density $%
P_{1}(x),$ Eq. (\ref{Discreta}). Hence, independently of the previous
history, it assumes the value $x_{\mu }$ with probability $q_{\mu }.$
Alternatively, with probabilities $n_{\mu }/(n+\lambda ),$ which depends on
all previous history, it assumes the value $x_{\mu }.$ The parameter $%
\lambda $ measure the weigh of both options.

In order to achieve a deeper understanding of the processes defined by Eq. (%
\ref{TransitionDirichlet}), in Fig. 1 we plotted a set of realizations for
the random variables$\ \{X_{i}\}_{i=1}^{n}$ (upper panels). They correspond
to classical spin variables, that is, we take $x_{\mu }=\pm 1$ and $M=2.$
For clarity, each value of $X_{i}$ is continued in the real interval $%
(i-1,i).$

The first value, $X_{1},$ is chosen in agreement with $P_{1}(x),$ Eq. (\ref%
{Discreta}). The next values $\{X_{i}\}_{i=2}^{n}$ follows from the
transition probability $T_{n}(\{x_{i}\}|x),$ Eq. (\ref{TransitionDirichlet}%
). We also plotted this object as a function of $n$ and for each value of $%
x=\pm 1$ (lower panels). Notice that each curve gives the probability for
the next variable, given the previous history. Therefore they are random
objects. We found that for increasing $n,$ the transition probabilities
always saturate to stationary values. Therefore, when this regime is
achieved, each realization is equivalent to that of independent random
variables. Nevertheless, the stationary values reached by the transition
probabilities are different for each realization, that is, they are random.
This property, valid for any $\lambda ,$ is characterized in the next
Section [see Eq. (\ref{Pu})].

In the realization of Fig. 1(a) the stationary transition probability for
the state $-1$ is larger than for the state $+1.$ Consistently, the state $%
-1 $ is taken much more frequently, feature clearly visible in the upper
panel. In Fig. 1(b) the difference between the stationary values is much
smaller, inducing a more \textquotedblleft noisy\textquotedblright\
realization.

In Fig. 2 we plot a set of realizations obtained from the transition
probability Eq. (\ref{TransitionDirichlet}) for random variables
characterized by three states, $M=3,$ with $x_{\mu }=+1,0,-1.$ Similarly to
the case of two-level variables, for increasing $n$ the transition
probabilities reach stationary values, which are different and random for
each realization. Therefore, in this regime the realizations are also
equivalent to that of identical independent random variables. In Fig. 2(a)
the random variables almost always assume the values $x=\pm 1.$ This happens
because the stationary value of the transition probability corresponding the
state $x=0$ is much smaller than the other two, $x=\pm 1.$ Instead, in Fig.
2(b) the state $x=0$ has the larger stationary transition probability.
Consistently, the states $x=\pm 1$ appear sparsely.%
\begin{figure}[tbp]
\includegraphics[bb=60 836 687 1103,angle=0,width=8.6cm]{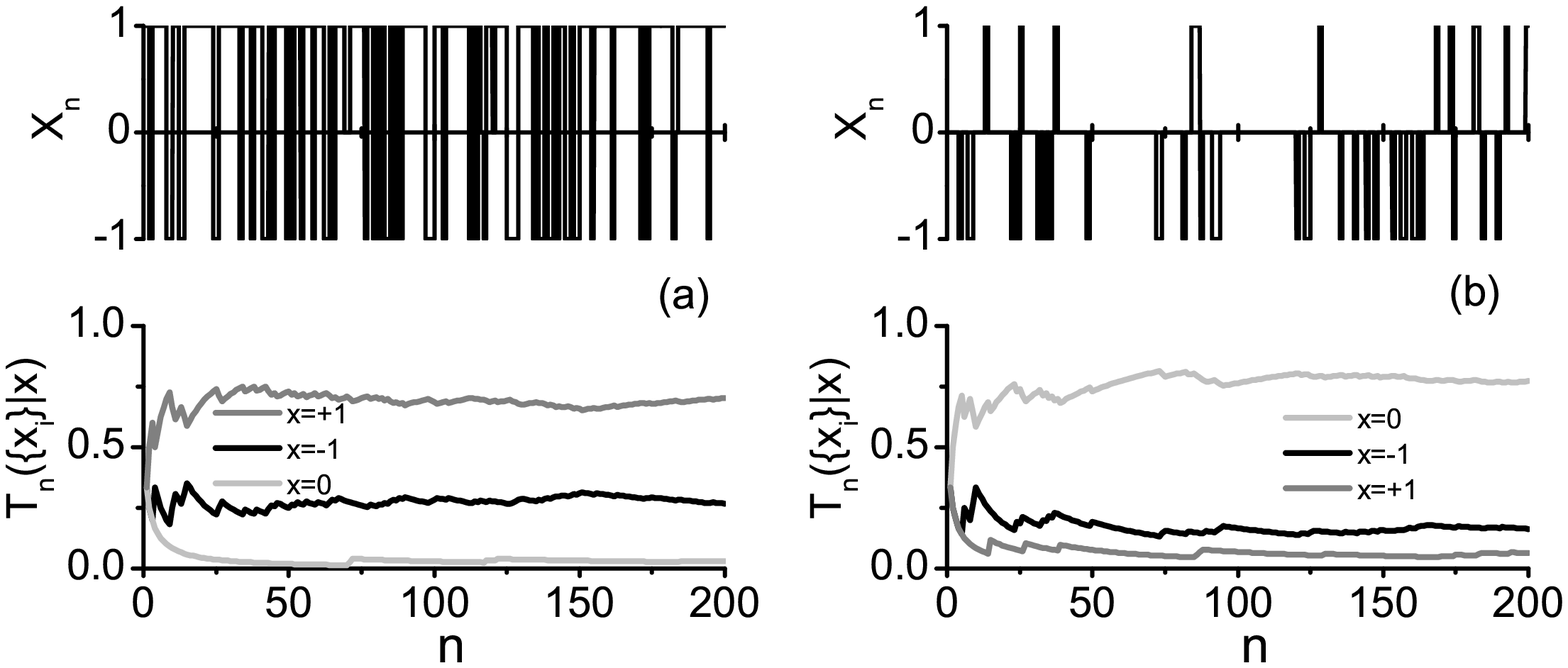}
\caption{Two realizations [(a) and (b)] for a set of three-state random
variables $\{x_{\protect\mu }\}=\{+1,0,-1\}.$ They follows from the
transition probability (\protect\ref{TransitionDirichlet}) $(M=3),$ which is
plotted in the lower panels. The parameters are $q_{+}=q_{-}=q_{0}=1/3$ and $%
\protect\lambda =3.$}
\end{figure}

\subsection*{Composed P\'{o}lya Urn scheme}

Here, we introduce a generalization of the previous urn scheme that is also
consistent with interchangeability. We consider non-discrete random
variables with \textit{arbitrary} probability density $P_{1}(x).$ The domain 
$\Omega $ of each variable $\{X_{i}\}_{i=1}^{n},$ that is, the domain of $%
P_{1}(x),$ is split in a finite set of disjoint subdomains $\{\Omega _{\mu
}\}_{\mu =1}^{M}$ such that the total domain is their union, $\Omega =\cup
\Omega _{\mu }.$ To each region $\Omega _{\mu }$ we associate\ a probability
density $p_{\mu }(x),$ normalized as $\int_{\Omega }p_{\mu }(x)dx=1.$ Under
these definitions, we propose the transition probability density%
\begin{equation}
T_{n}(\{x_{i}\}|x)=\frac{\lambda P_{1}(x)+\sum\limits_{\mu =1}^{M}p_{\mu
}(x)\sum\limits_{i=1}^{n}\int_{\Omega _{\mu }}dy\delta (y-x_{i})}{n+\lambda }%
.  \label{TransitionEstructurado}
\end{equation}%
As before, $\lambda $ is a free parameter and $\{x_{i}\}=x_{1},x_{2},\cdots
x_{n}$ is the previous trajectory.

The integral contributions%
\begin{equation}
n_{\mu }\equiv \sum_{i=1}^{n}\int_{\Omega _{\mu }}dy\delta (y-x_{i}),\ \ \ \
\ \ \ \ \sum_{\mu =1}^{M}n_{\mu }=n,  \label{numerosENE}
\end{equation}%
give the number of times the variables $\{x_{i}\}_{i=1}^{n}$ fell in the
subdomain $\Omega _{\mu }.$ Therefore, we can write%
\begin{equation}
T_{n}(\{x_{i}\}|x)=\frac{\lambda P_{1}(x)+\sum_{\mu =1}^{M}p_{\mu }(x)n_{\mu
}}{n+\lambda }.  \label{numeros}
\end{equation}%
The correlation dynamics induced by Eq. (\ref{TransitionEstructurado}) is
then clear. With probability $\lambda /(n+\lambda )$ the next variable,
independently of the previous history, is chosen in agreement with $%
P_{1}(x). $ On the other hand, with probabilities $n_{\mu }/(n+\lambda ),$
the next value is chosen in agreement with the arbitrary densities $p_{\mu
}(x).$

It is simple to check that the transition probability density (\ref%
{TransitionEstructurado}), for arbitrary domains $\{\Omega _{\mu }\}_{\mu
=1}^{M}$ and densities $\{p_{\mu }(x)\}_{\mu =1}^{M},$ is normalized and
positive defined%
\begin{equation}
\int_{\Omega }dxT_{n}(\{x_{i}\}|x)=1,\ \ \ \ \ \ \ \ \ \
T_{n}(\{x_{i}\}|x)\geq 0.
\end{equation}%
On the other hand, in order to be consistent with the interchangeability
symmetry it must to satisfy the hierarchical relations Eq. (\ref%
{RecursivaCondicional}). After same algebra, it follows that
interchangeability is fulfilled under the condition%
\begin{equation}
\sum_{\mu =1}^{M}p_{\mu }(x)\int_{\Omega _{\mu }}P_{1}(y)dy=P_{1}(x),
\label{C1}
\end{equation}%
jointly with the following one,%
\begin{equation}
\int_{\Omega _{\mu }}p_{\mu ^{\prime }}(x)dx=\delta _{\mu \mu ^{\prime }}.
\label{C2}
\end{equation}%
Hence, interchangeability is not fulfilled by arbitrary densities $\{p_{\mu
}(x)\}.$

Condition (\ref{C1}) say us that the set $\{p_{\mu }(x)\},$ under
appropriate weights, recover the distribution $P_{1}(x).$ Condition (\ref{C2}%
) implies that each density $p_{\mu }(x)$\ is not null only on its
associated subdomain $\Omega _{\mu }.$ A solution to these constraints is
given by%
\begin{equation}
p_{\mu }(x)=P_{1}(x)\frac{\theta _{\Omega _{\mu }}(x)}{\int_{\Omega _{\mu
}}P_{1}(x^{\prime })dx^{\prime }},  \label{pOmega}
\end{equation}%
where we defined the region indicator%
\begin{equation}
\theta _{\Omega _{\mu }}(x)\equiv \left\{ 
\begin{array}{c}
1\ \ if\ x\in \Omega _{\mu } \\ 
0\ \ if\ x\notin \Omega _{\mu }%
\end{array}%
\right. .  \label{tetal}
\end{equation}%
It is simple to check that (\ref{pOmega}) satisfies both constraints. Hence,
interchangeability is fulfilled.

Interestingly, from the previous solutions for $\{p_{\mu }(x)\}_{\mu
=1}^{M}, $ Eq. (\ref{pOmega}), we can write the probability density of each
variable as%
\begin{equation}
P_{1}(x)=\sum_{\mu =1}^{M}q_{\mu }p_{\mu }(x),  \label{CF1}
\end{equation}%
where the positive weights are%
\begin{equation}
q_{\mu }\equiv \int_{\Omega _{\mu }}P_{1}(x)dx,  \label{CF2}
\end{equation}%
and fulfill $\sum\nolimits_{\mu =1}^{M}q_{\mu }=1.$ In consequence, by using
Eq. (\ref{numeros}), the transition probability Eq. (\ref%
{TransitionEstructurado}) reads%
\begin{equation}
T_{n}(\{x_{i}\}|x)=\sum_{\mu =1}^{M}\frac{\lambda q_{\mu }+n_{\mu }}{%
n+\lambda }p_{\mu }(x).  \label{composed}
\end{equation}%
This final expression is the main result of this section.

Eq. (\ref{composed}) has a stretched relation with the standard P\'{o}lya
urn scheme, Eq. (\ref{TransitionDirichlet}). In fact, both expressions are
related by the replacements $\delta (x-x_{\mu })\leftrightarrow p_{\mu }(x).$
On the other hand, the weights in the single density Eq. (\ref{Discreta})
here follows from Eqs. (\ref{CF1}) and (\ref{CF2}). Hence, each subdomain $%
\Omega _{\mu }$ can be associated to the states $x_{\mu }$ [Eq. (\ref%
{Discreta})]. Nevertheless, instead of the value $x_{\mu },$ here the next
variable assumes a random value distributed over the subdomain $\Omega _{\mu
}$ with probability density $p_{\mu }(x).$ In consequence, Eq. (\ref%
{composed}) can be read as an independent statistical \textit{composition}
of the P\'{o}lya urn scheme, Eq. (\ref{TransitionDirichlet}), and the set of
probability densities $\{p_{\mu }(x)\}_{\mu =1}^{M}.$ %
\begin{figure}[tbp]
\includegraphics[bb=60 836 687 1103,angle=0,width=8.6cm]{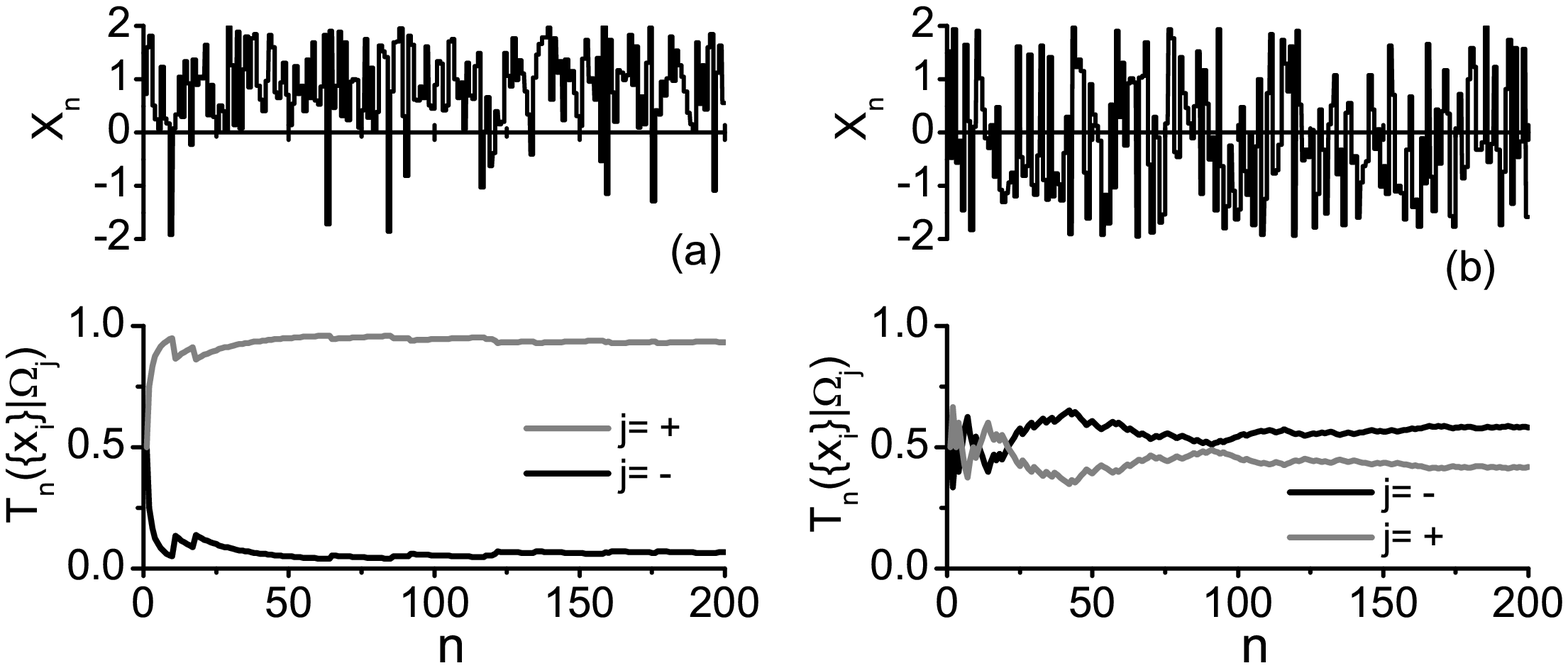}
\caption{Two realizations [(a) and (b)] of the composed P\'{o}lya urn scheme
[Eq. (\protect\ref{composed})] defined by the Eqs. (\protect\ref{urn1}) and (%
\protect\ref{urn2}). The lower panels correspond to the transition
probabilities associated to the subdomains $\Omega _{+}$ and $\Omega _{-},$
each one having weights $q_{+}=q_{-}=1/2$ (see text). The parameter is $%
\protect\lambda =2.$}
\end{figure}

As an example, we take the uniform probability density%
\begin{equation}
P_{1}(x)=\frac{1}{4},\ \ \ \ \ \ \ \ -2\leq x\leq 2,  \label{urn1}
\end{equation}%
and $P_{1}(x)=0$ if $x\notin \lbrack -2,2].$ Therefore, each variable only
assume random values over the real interval $[-2,2].$ Then $\Omega =\{x\in
\lbrack -2,2]\}.$ The composed urn scheme is completely characterized after
defining the subdomains $\{\Omega _{\mu }\}.$ We consider only two
subspaces, $\Omega _{+}$ and $\Omega _{-},$ defined as $\Omega _{+}=\{x\in
\lbrack 0,2]\}$ and $\Omega _{-}=\{x\in \lbrack -2,0)\}$ respectively.
Notice that $\Omega =\Omega _{+}\cup \Omega _{-}.$The associated probability
densities, from Eqs. (\ref{pOmega}) and (\ref{tetal}) becomes 
\begin{subequations}
\label{urn2}
\begin{eqnarray}
p_{+}(x) &=&\frac{1}{2},\ \ \ \ \ \ \ \ 0\leq x\leq 2, \\
p_{-}(x) &=&\frac{1}{2},\ \ \ \ \ \ \ \ -2\leq x<0.
\end{eqnarray}%
Notice that the underlying discrete process that decides which probability
density is chosen, $p_{+}(x)$ or $p_{-}(x),$ is equivalent to that plotted
in Fig. 1. In fact, from Eq. (\ref{CF2}) if follows $q_{+}=q_{-}=1/2.$

In Fig. 3 we plot a set of realizations corresponding to the previous
definitions. In contrast to the previous figures, here the random variables
assume values over the real interval $[-2,2].$ In the lower panels we plot
the underlying transition probability $T_{n}(\{x_{i}\}|\Omega _{j})$ with
governs which subdomain $(j=\pm )$ is occupied in the next step.
Consistently, its behavior is similar to that of Fig. 1. In Fig. 3(a) the
subspace $\Omega _{+}$\ has a higher stationary probability and,
consistently, the realization take most of its values in the interval $%
[0,2]. $ In Fig. 3(b) both subspaces have similar stationary values. Hence,
the realization looks like a random signal in the full domain $[-2,2].$ On
the other hand, by averaging over realizations we checked that the
probability density of each variable $\{X_{i}\}_{i=1}^{n}$ is given by $%
P_{1}(x),$ Eq. (\ref{urn1}).

\section{Statistics of the sum variable}

In the previous Section (and in the Appendixes) we described different
memory mechanism and statistics consistent with the interchangeability
property. Here, we study departures with respect to the standard CLT when
considering such kind of globally correlated variables.

We consider the normalized random sum variable 
\end{subequations}
\begin{equation}
W=\lim_{n\rightarrow \infty }W_{n}=\lim_{n\rightarrow \infty }\Big{(}\frac{1%
}{n}\sum_{i=1}^{n}X_{i}\Big{)}.  \label{Wdefinition}
\end{equation}%
Notice that in contrast with the standard CLT \cite%
{kolmogorov,feller,vanKampen,gardiner}, instead of $1/\sqrt{n},$ here the
normalization is $1/n.$ We choose this factor because all studied models,
depending on their characteristic parameters, are able to reach a full
correlated regime where all variables $\{X_{i}\}$ assume the same random
value. Hence, in that regime the normalization $1/n$ is the only one that
delivers a random variable $(W)$ that (asymptotically) does not depend on $%
n. $

The probability density $P(w)$ of $W$ can be written as the following limit, 
$P(w)=\lim_{n\rightarrow \infty }P(w_{n}),$%
\begin{equation*}
P(w)=\lim_{n\rightarrow \infty }\int dx_{1}\cdots dx_{n}\delta \Big{(}w-%
\frac{1}{n}\sum_{i=1}^{n}x_{i}\Big{)}P_{n}(\{x_{i}\}),
\end{equation*}%
where $P_{n}(\{x_{i}\})\equiv P_{n}(x_{1},\cdots x_{n})$ is the $n$-joint
probability density. By introducing the Fourier representation of the delta
Dirac function, $\delta (x)=(1/2\pi )\int_{-\infty }^{+\infty }e^{-ikx}dk,$
the characteristic function $G_{w}(k)$ of $P(w),$%
\begin{equation}
G_{w}(k)=\int_{-\infty }^{+\infty }dwe^{ikw}P(w),  \label{caracteristica}
\end{equation}%
can be written as, $G_{w}(k)=\lim_{n\rightarrow \infty }G_{w_{n}}(k),$%
\begin{equation}
G_{w}(k)=\lim_{n\rightarrow \infty }\int dx_{1}\cdots dx_{n}\exp \Big{(}i%
\frac{k}{n}\sum_{i=1}^{n}x_{i}\Big{)}P_{n}(\{x_{i}\}).  \label{FourierSuma}
\end{equation}%
In terms of the the multiple Fourier transform of $P_{n}(\{x_{i}\}),$ that
is,%
\begin{equation*}
G_{n}(\{k_{i}\})=\int dx_{1}\cdots dx_{n}\exp \Big{(}%
i\sum_{i=1}^{n}k_{i}x_{i}\Big{)}P_{n}(\{x_{i}\}),
\end{equation*}%
it follows%
\begin{equation}
G_{w}(k)=\lim_{n\rightarrow \infty }G_{w_{n}}(k)=\lim_{n\rightarrow \infty
}G_{n}(\{k_{i}=\frac{k}{n}\}).  \label{GwFinal}
\end{equation}

Below we treat the different cases introduced previously. For clarifying the
derivation of some results, the well known case of independent variables is
reviewed first.

\subsection{Statistical independent variables}

Assume the set $\{X_{i}\}_{i=1}^{n}$ are independent random variables with
probability density distribution $P_{1}(x).$ Therefore, $P_{n}(\{x_{i}\})=%
\prod_{i=1}^{n}P_{1}(x_{i}).$ From Eq. (\ref{FourierSuma}), it follows%
\begin{equation}
G_{w_{n}}(k)=\Big{[}G_{x}\Big{(}\frac{k}{n}\Big{)}\Big{]}^{n},
\label{GIndependiente}
\end{equation}%
where $G_{x}(k)$ is the Fourier transform of $P_{1}(x).$

For small $k,$ we assume valid the approximation $G_{x}(k)\simeq
e^{ik\left\langle x\right\rangle }(1-\frac{\sigma ^{2}k^{2}}{2}),$ where 
\begin{equation}
\bar{x}=\int_{-\infty }^{+\infty }dxxP_{1}(x),\ \ \ \ \ \ \ \sigma
^{2}=\int_{-\infty }^{+\infty }dx(x-\bar{x})^{2}P_{1}(x),  \label{momento}
\end{equation}%
are the mean value and standard deviation of the distribution $P_{1}(x).$
Therefore, we can approximate $G_{w_{n}}(k)\simeq e^{ik\bar{x}}[1-\frac{%
\sigma ^{2}k^{2}}{2n^{2}}]^{n},$ which can be rewritten as%
\begin{equation}
G_{w_{n}}(k)\simeq e^{ik\bar{x}}\exp \Big{[}-\frac{1}{2}\frac{\sigma
^{2}k^{2}}{n}\Big{]}.  \label{LawLN}
\end{equation}%
After Fourier inversion it follows%
\begin{equation}
P(w_{n})\simeq \sqrt{\frac{1}{2\pi (\sigma ^{2}/n)}}\exp \Big{[}-\frac{1}{2}%
\frac{(w_{n}-\bar{x})^{n}}{(\sigma ^{2}/n)}\Big{]},
\label{GaussianaIndependiente}
\end{equation}%
which is a Gaussian distribution. Given that $P(w)=\lim_{n\rightarrow \infty
}P(w_{n}),$ it follows that%
\begin{equation}
P(w)=\delta (w-\bar{x}).  \label{DeltaIndependientes}
\end{equation}%
Therefore, the random variable $W$ deterministically assume the value $\bar{x%
}.$ This result, which can be read as the well known law of large numbers 
\cite{kolmogorov,feller,vanKampen,gardiner}, follows from the normalization $%
1/n$ in Eq. (\ref{Wdefinition}). In fact, defining the variable $\sqrt{n}W$
from Eq. (\ref{GaussianaIndependiente}) one recovers a Gaussian distribution
that does not depends on $n,$ which in turn corresponds to the standard CLT.
The basin or domain of attraction of the normal distribution corresponds to
all distributions $P_{1}(x)$ with finite first and second moments.

Using the same Fourier techniques, we showed that departure with respect to
Eq. (\ref{DeltaIndependientes}) arise from (correlated) Gaussian variables
[see Eq. (\ref{PGaussol})]\ and also in the de Finetti representation [see
Eq. (\ref{PSumaFinetti})]. In fact, the possibility of achieving a fully
correlated regime is enough for warranting departure from a delta
distribution.

\subsection{P\'{o}lya Urn scheme}

In the previous section, we explicitly showed a very important property of
the P\'{o}lya urn scheme, that is, for increasing $n$ the transition
probabilities converge to that of identical independent random variables.
Nevertheless, the stationary values achieved by the transition probability
are random, that is, their are different for each realization. This property
was characterized previously in the literature \cite{queen,pitman}. Here, we
review it in order to characterize the sum variable (\ref{Wdefinition}).

The transition probability Eq. (\ref{TransitionDirichlet}), in the
asymptotic regime is characterized by the weights%
\begin{equation}
F_{\mu }\equiv \lim_{n\rightarrow \infty }\frac{\lambda q_{\mu }+n_{\mu }}{%
n+\lambda },\ \ \ \ \ 0\leq F_{\mu }\leq 1,  \label{Pu}
\end{equation}%
which consistently satisfy $\sum\nolimits_{\mu =1}^{M}F_{\mu }=1.$ These
weights (probabilities) are different for each realization, that is, their
are random variables. Hence, taking an ensemble of realizations [see Figs.
(1) and (2)] one can define their probability density $D(\{f_{\mu
}\}|\{\lambda _{\mu }\}),$ which depends on the characteristic parameters of
the problem, here defined as%
\begin{equation}
\lambda _{\mu }\equiv \lambda q_{\mu }.
\end{equation}%
Due to the normalization of the weights $\{q_{\mu }\}_{\mu =1}^{M},$ it
follows $\lambda =\sum\nolimits_{\mu =1}^{M}\lambda _{\mu }.$ It is known
that $D(\{f_{\mu }\}|\{\lambda _{\mu }\})$ is a \textit{Dirichlet
distribution} \cite{queen,pitman},%
\begin{equation}
D(\{f_{\mu }\}|\{\lambda _{\mu }\})\equiv \frac{\Gamma (\lambda )}{%
\prod_{\mu ^{\prime }=1}^{M}\Gamma (\lambda _{\mu ^{\prime }})}\prod_{\mu
=1}^{M}f_{\mu }^{\lambda _{\mu }-1},  \label{DirichletDensity}
\end{equation}%
where $\Gamma (x)$\ is the Gamma function. $D(\{f_{\mu }\}|\{\lambda _{\mu
}\})$ is positive for all values of $\{f_{\mu }\}_{\mu =1}^{M},$ and
normalized as $\int_{\Lambda }df_{1}\cdots df_{M-1}\ D(\{f_{\mu
}\}|\{\lambda _{\mu }\})=1,$ where $\Lambda $ is the region defined by $%
\sum\nolimits_{\mu =1}^{M}f_{\mu }=1.$ On the other hand, the relation $%
q_{\nu }=\int_{\Lambda }df_{1}\cdots df_{M-1}\ D(\{f_{\mu }\}|\{\lambda
_{\mu }\})f_{\nu }$ is fulfilled for all $\nu =1,\cdots M.$ 
\begin{figure}[tbp]
\includegraphics[bb=55 560 715 1110,angle=0,width=9cm]{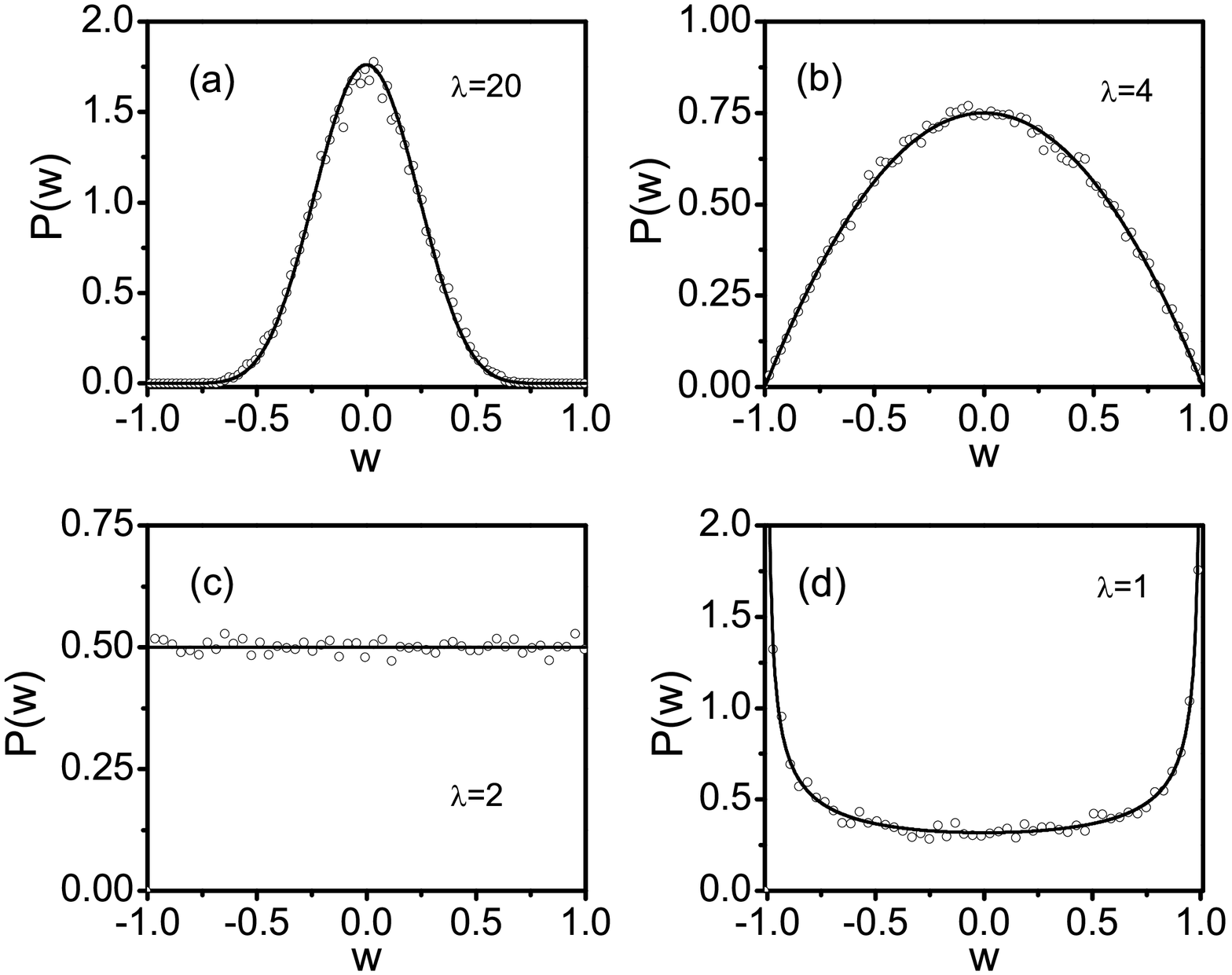}
\caption{Probability density of the sum variable $W,$ Eq. (\protect\ref%
{Wdefinition}). The random contributions are classical spin variables, $x_{%
\protect\mu }=\pm 1,$ defined by the\ transtion probability (\protect\ref%
{TransitionDirichlet}). The full lines correspond to the analytical
expression Eq. (\protect\ref{SpinPw}), while the circles correspond to
numerical simulations with $n=300$ terms and $5\times 10^{5}$ realizations
(see Fig. 1). In all cases, $q_{+}=q_{-}=1/2.$}
\end{figure}

From Eq. (\ref{DirichletDensity}) we can obtain the probability density $P(w)
$\ of the variable $W.$ Given that asymptotically each realization is
equivalent to that of independent random variables, one can associate the
probability density $\delta (w-\bar{X}_{f})$ to each realization [see Eq. (%
\ref{DeltaIndependientes})], where $\bar{X}_{f}=\sum\nolimits_{\mu
=1}^{M}F_{\mu }x_{\mu }.$ Now, the final structure of $P(w)$ arises after
averaging over realizations. Given that the random variables $F_{\mu }$
obeys the statistics given by Eq. (\ref{DirichletDensity}), it follows%
\begin{equation}
P(w)=\int_{\Lambda }df_{1}\cdots df_{M-1}\ \delta (w-\bar{x}_{f})D(\{f_{\mu
}\}|\{\lambda _{\mu }\}),  \label{PwPolya}
\end{equation}%
where%
\begin{equation}
\bar{x}_{f}\equiv \sum\nolimits_{\mu =1}^{M}f_{\mu }x_{\mu }.
\label{xpPolya}
\end{equation}

From the result Eq. (\ref{PwPolya}), in the limit of $\lambda \rightarrow
\infty $ consistently we recover the independent random variables case, $%
\lim_{\lambda \rightarrow \infty }P(w)=\delta (w-\bar{x}_{q}),$ where $\bar{x%
}_{q}\equiv \sum\nolimits_{\mu =1}^{M}q_{\mu }x_{\mu }.$ In the limit $%
\lambda \rightarrow 0,$ which corresponds to the fully correlated case, it
follows $\lim_{\lambda \rightarrow 0}P(w)=\sum\nolimits_{\mu =1}^{M}q_{\mu
}\delta (w-x_{\mu })$ [see Eq. (\ref{TransitionDirichlet})].

The final expression (\ref{PwPolya}) allow us to characterize the CLT for
the P\'{o}lya urn scheme. It is valid for any value of $M$ and arbitrary
discrete distributions, Eq. (\ref{Discreta}). For example, for classical
spin variables, $x_{\mu }=\pm 1,$ after integration we get $(\lambda _{\pm
}\equiv \lambda q_{\pm })$%
\begin{equation}
P(w)=\frac{1}{\mathcal{N}}(1+w)^{\lambda _{+}-1}(1-w)^{\lambda _{-}-1},
\label{SpinPw}
\end{equation}%
where $\mathcal{N}\equiv 2^{\lambda _{+}+\lambda _{-}-1}\Gamma (\lambda
_{+})\Gamma (\lambda _{-})/\Gamma (\lambda _{+}+\lambda _{-}).$

We notice that Eq. (\ref{SpinPw}) can be read as a Beta \cite{feller} or
asymmetric $q$-Gaussian distribution \cite{budini}. In the symmetric case $%
\lambda _{+}=\lambda _{-},$ this result was derived previously in the
context of a nonextensive thermodynamics approach \cite{Qhanel} (see also 
\cite{feller,norman}).

In Fig. 4 we obtained numerically $P(w)$ by averaging a set of realizations
such as those presented in Fig. 1. Results for different values of $\lambda $%
\ are presented, while $q_{+}=q_{-}=1/2.$ Independently of the parameter
values, we find that Eq. (\ref{SpinPw}) fits the numerical results. 
\begin{figure}[tbp]
\includegraphics[bb=55 560 715 1110,angle=0,width=9cm]{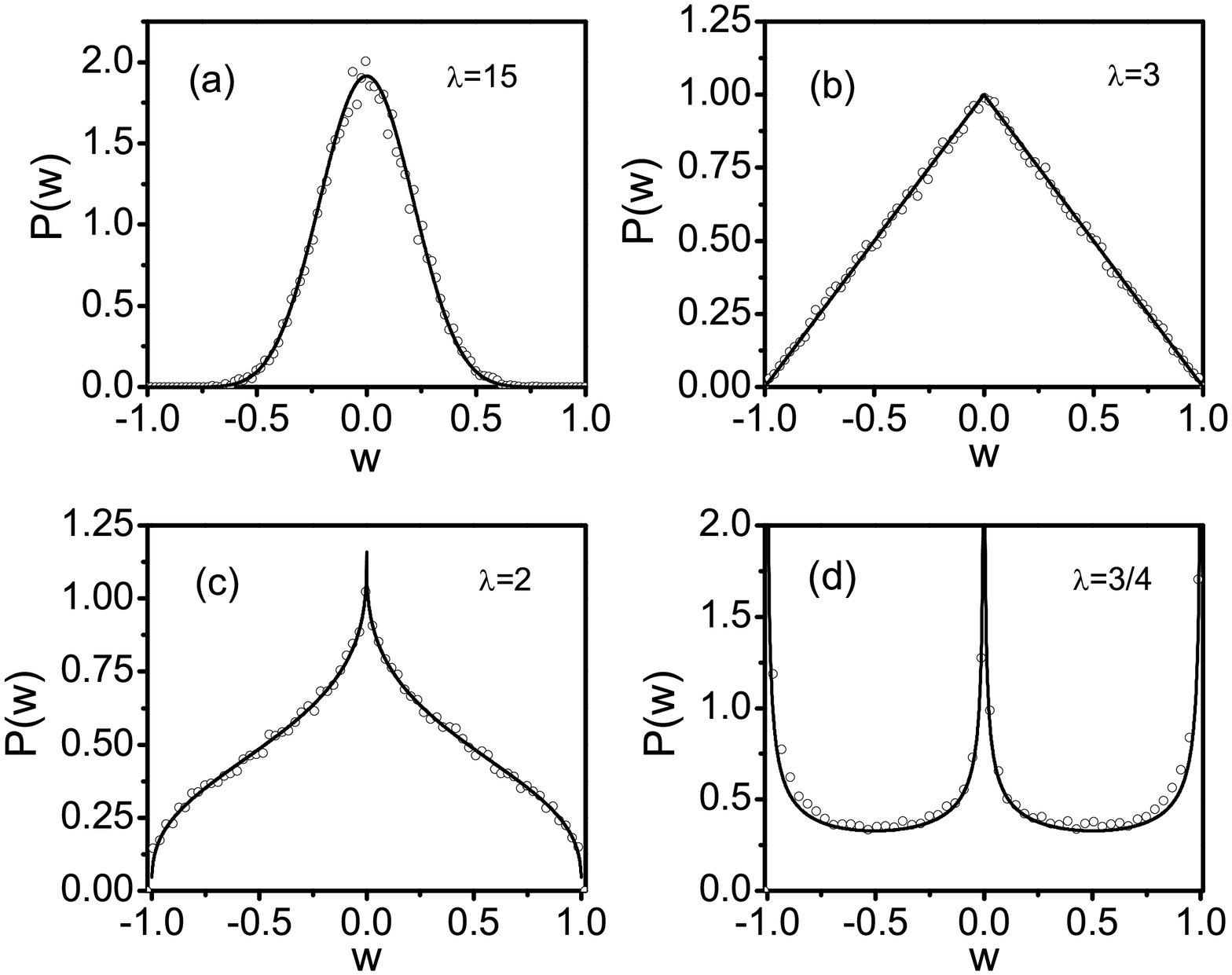}
\caption{Probability density $P(w)$ for random variables with three discrete
states, $x_{\protect\mu }=+1,0,-1,$ obtained from the\ transtion probability
(\protect\ref{TransitionDirichlet}). The full lines correspond to the
analytical expression Eq. (\protect\ref{PwTres}), while the circles
correspond to a numerical simulation with $n=300$ terms and $5\times 10^{5}$
realizations (see Fig. 2). In all cases, $q_{+}=q_{0}=q_{-}=1/3.$}
\end{figure}

For three-states variables with $\{x_{\mu }\}=\{+1,0,-1\},$ the parameters
are $\{q_{\mu }\}=\{q_{+},q_{0},q_{-}\}$ and $\lambda .$ They can be
parametrized as $\{\lambda _{\mu }\}=\{\lambda q_{\mu }\}=\{\lambda
_{+},\lambda _{0},\lambda _{-}\}.$ By taking into account that $\bar{x}%
_{f}=f_{+}-f_{-}$ [Eq. (\ref{xpPolya})], from Eq. (\ref{PwPolya}) we get,%
\begin{equation}
P(w)=\left\{ 
\begin{array}{l}
g_{+}[w]\ \ \ \ \ \ \ \ \ w>0 \\ 
g_{-}[w]\ \ \ \ \ \ \ \ \ w<0%
\end{array}%
\right. ,  \label{PwTres}
\end{equation}%
where each contribution is defined as%
\begin{eqnarray*}
g_{+}[w] &=&\int_{0}^{\frac{1-w}{2}}df\ f^{\lambda _{-}-1}(1-w-2f)^{\lambda
_{0}-1}(f+w)^{\lambda _{+}-1}, \\
g_{-}[w] &=&\int_{0}^{\frac{1+w}{2}}df\ (f-w)^{\lambda
_{-}-1}(1+w-2f)^{\lambda _{0}-1}f^{\lambda _{+}-1}.
\end{eqnarray*}%
These integrals can be solved in terms of the hypergeometric function $%
_{2}F_{1}[a,b,c,z]$ as%
\begin{eqnarray*}
g_{+}[w] &=&\frac{(1-w)^{\lambda _{-}+\lambda _{0}-1}w^{\lambda _{+}-1}}{%
2^{\lambda _{-}}\Gamma ^{-1}(\lambda )\Gamma (\lambda _{+})\Gamma (\lambda
_{0}+\lambda _{-})} \\
&&_{2}F_{1}[\lambda _{-},1-\lambda _{+},\lambda _{0}+\lambda _{-},\frac{w-1}{%
2w}],\ 
\end{eqnarray*}%
and similarly%
\begin{eqnarray*}
g_{-}[w] &=&\frac{(1+w)^{\lambda _{+}+\lambda _{0}-1}(-w)^{\lambda _{-}-1}}{%
2^{\lambda _{+}}\Gamma ^{-1}(\lambda )\Gamma (\lambda _{-})\Gamma (\lambda
_{0}+\lambda _{+})} \\
&&_{2}F_{1}[1-\lambda _{-},\lambda _{+},\lambda _{0}+\lambda _{+},\frac{w+1}{%
2w}].\ 
\end{eqnarray*}%
The hypergeometric function is defined by $_{2}F_{1}[a,b,c,z]=\sum_{k=0}^{%
\infty }(a)_{k}(b)_{k}(c)_{k}z^{k}/k!$ with $(x)_{k}=\prod_{j=0}^{k}(x+j).$
Simpler expressions can be found in the particular case $\lambda
_{+}=\lambda _{-1}=1,$ where Eq. (\ref{PwTres}) reduces to%
\begin{equation}
P(w)=\frac{1}{2}(1+\lambda _{0})(1-|w|)^{\lambda _{0}},\ \ \ \ \ \ \ \ \
\lambda _{+}=\lambda _{-1}=1.  \label{PTresSimple}
\end{equation}

In Fig. 5 we show a set of plots corresponding to $P(w).$ The realizations,
over which the distribution are obtained, are those shown in Fig. 2. We
found that the density (\ref{PwTres}) fits the numerical results. The case
shown Fig. 5(b) corresponds to Eq. (\ref{PTresSimple}) with $\lambda _{0}=1.$

\subsection{Composed P\'{o}lya Urn scheme}

The previous results with the P\'{o}lya urn scheme (see also the Appendixes)
demonstrates that the sum variable Eq. (\ref{Wdefinition}), depending on the
underlying correlation mechanism, may adopt very different statistics. In
contrast to independent random variables, these probabilities do not have
associated a basin of attraction. Here, we show that the composed P\'{o}lya
urn scheme fall in the basin of attraction of the standard scheme. This is
the main result of this section.

For the composed P\'{o}lya urn scheme, the probability density of the sum
variable $W$ [Eq. (\ref{Wdefinition})] is given by Eq. (\ref{PwPolya}) with $%
\{q_{\mu }\}_{\mu =1}^{M}$ given by Eq. (\ref{CF2}) and under the replacement%
\begin{equation}
\bar{x}_{f}=\sum\nolimits_{\mu =1}^{M}f_{\mu }\bar{x}_{\mu },\ \ \ \ \ \ \ \
\ \ \ \bar{x}_{\mu }=\int_{\Omega _{\mu }}dxxp_{\mu }(x).
\label{AveragesComposed}
\end{equation}%
Therefore, the main change corresponds to $x_{\mu }\rightarrow \bar{x}_{\mu
} $ [Eq. (\ref{xpPolya})] where $\bar{x}_{\mu }$ is the mean value
associated to the distribution $p_{\mu }(x).$ This result say us that all
random variables obtained from the composed P\'{o}lya urn scheme are in the
basin of the attractors corresponding to the standard urn scheme, Eq. (\ref%
{PwPolya}). As shown below, this result relies on the applicability of the
law of large numbers to random independent variables draw randomly from any
of the distributions $\{p_{\mu }(x)\}_{\mu =1}^{M}.$

For demonstrating the previous result we use that the composed P\'{o}lya urn
scheme consist of two independent random processes: the randomness
introduced by the probability densities $p_{\mu }(x)$ associated to each
subdomain $\Omega _{\mu }$ and the underlying P\'{o}lya urn process that
select each subdomain. Therefore, the joint probability density of the
random variables $\{X_{i}\}_{i=1}^{n}$ reads%
\begin{equation}
P_{n}(\{x_{i}\})=\left\langle p_{\mu _{1}}(x_{1})\cdots p_{\mu
_{n}}(x_{n})\right\rangle _{\{\mu \}}.
\end{equation}%
Here, each index $\mu _{i}=1\cdots M$ runs over the set of subdomains $%
\{\Omega _{\mu }\}_{\mu =1}^{M}.$ On other hand, $\left\langle \cdots
\right\rangle _{\{\mu \}}$ denotes and average over the ensemble of
realizations associated \ to the underlying P\'{o}lya urn scheme. From Eqs. (%
\ref{FourierSuma}) and (\ref{GwFinal})\ we get $G_{w}(k)=\lim_{n\rightarrow
\infty }G_{w}^{(n)}(k),$ with%
\begin{equation}
G_{w}^{(n)}(k)=\left\langle G_{\mu _{1}}\Big{(}\frac{k}{n}\Big{)}\cdots
G_{\mu _{n}}\Big{(}\frac{k}{n}\Big{)}\right\rangle _{\{\mu \}},
\end{equation}%
where $G_{\mu }(k)$ is the Fourier transform of $p_{\mu }(x).$ By indexing
the realizations by the number of times $n_{\mu }$ that each subspace $%
\Omega _{\mu }$ is selected, we can write%
\begin{equation}
G_{w}^{(n)}(k)=\left\langle \Big{[}G_{1}\Big{(}\frac{k}{n}\Big{)}\Big{]}%
^{n_{1}}\cdots \Big{[}G_{M}\Big{(}\frac{k}{n}\Big{)}\Big{]}%
^{n_{M}}\right\rangle _{\{n\}}.  \label{NCaracteristica}
\end{equation}%
Taking into account that when $n\rightarrow \infty $ the set of occurrences
also diverge, $\{n_{\mu }\}\rightarrow \infty ,$ each factor in the previous
expression can be approximated as%
\begin{equation}
\Big{[}G_{\mu }\Big{(}\frac{k}{n}\Big{)}\Big{]}^{n_{\mu }}\approx \exp (ik%
\bar{x}_{\mu }\frac{n_{\mu }}{n})\exp (-\frac{\sigma _{\mu }^{2}k^{2}}{2n}%
\frac{n_{\mu }}{n}),  \label{hola}
\end{equation}%
where $\bar{x}_{\mu }$ is the mean value defined in Eq. (\ref%
{AveragesComposed}) while $\sigma _{\mu }^{2}=\int_{\Omega _{\mu }}dx(x-\bar{%
x}_{\mu })^{2}p_{\mu }(x).$ Notice that the previous approximation is
equivalent to the validity of the law of large numbers for each density $%
p_{\mu }(x)$ [see Eq. (\ref{LawLN})].

In the previous approximation, the argument $n_{u}/n,$ in the asymptotic
limit, can be associated with the random variables $F_{\mu },$ Eq. (\ref{Pu}%
). Therefore $n_{u}/n\simeq F_{\mu },$ which from Eqs. (\ref{NCaracteristica}%
) and (\ref{hola}) lead to%
\begin{equation}
G_{w}(k)=\left\langle \exp ik\sum_{\mu }F_{\mu }\bar{x}_{\mu }\right\rangle
_{\{F\}}.
\end{equation}%
The average over the random set of weights $\{F\}$ is governed by the
Dirichlet distribution Eq. (\ref{DirichletDensity}). Therefore, after
Fourier inversion we recover Eq. (\ref{PwPolya}), where instead of Eq. (\ref%
{xpPolya}), now it applies Eq. (\ref{AveragesComposed}). This finish the
demonstration.

As an example of the previous result we take a composed P\'{o}lya urn scheme
[Eq. (\ref{composed})] defined with two subdomains $\Omega _{\pm }$ with
densities $p_{\pm }(x).$ We get [Eq. (\ref{NCaracteristica})]%
\begin{equation}
G_{w}^{(n)}(k)=\left\langle \Big{[}G_{+}\Big{(}\frac{k}{n}\Big{)}\Big{]}^{n+}%
\Big{[}G_{-}\Big{(}\frac{k}{n}\Big{)}\Big{]}^{n_{-}}\right\rangle _{\{n\}},
\end{equation}%
where $n_{\pm }$ are the number of times that each subspace $\Omega _{\pm }$
was chosen, and $G_{\pm }(k)=\int_{-\infty }^{+\infty }dwe^{ikw}p_{\pm }(w).$
Using that $n_{+}+n_{-}=n,$ it follows%
\begin{equation}
G_{w}^{(n)}(k)=\sum_{n_{+}=0}^{n}P_{n}(n_{+})\Big{[}G_{+}\Big{(}\frac{k}{n}%
\Big{)}\Big{]}^{n_{+}}\Big{[}G_{-}\Big{(}\frac{k}{n}\Big{)}\Big{]}^{n-n_{+}},
\label{FourierSolution}
\end{equation}%
where $P_{n}(n_{+})$ is the probability of the random variable $n_{+}.$ This
object, after some algebra and by using the properties of Gamma functions,
can be obtained from from Eqs. (\ref{Conjunta}) and (\ref%
{TransitionDirichlet}). Alternatively, it can be obtained directly from de
Finetti representation theorem [see. Eq. (\ref{ProbNmas})]. It reads%
\begin{equation}
P_{n}(n_{+})=\frac{1}{\mathcal{N}_{n}}\binom{n}{n_{+}}\frac{\Gamma
(n_{+}+\lambda _{+})}{\Gamma (\lambda _{+})}\frac{\Gamma (n-n_{+}+\lambda
_{-})}{\Gamma (\lambda _{-})},
\end{equation}%
where $\mathcal{N}_{n}=\Gamma (n+\lambda )/\Gamma (\lambda ),$ and $\lambda
_{\pm }=\lambda q_{\pm }$ [Eq. (\ref{CF2})]. 
\begin{figure}[tbp]
\includegraphics[bb=55 560 715 1110,angle=0,width=9cm]{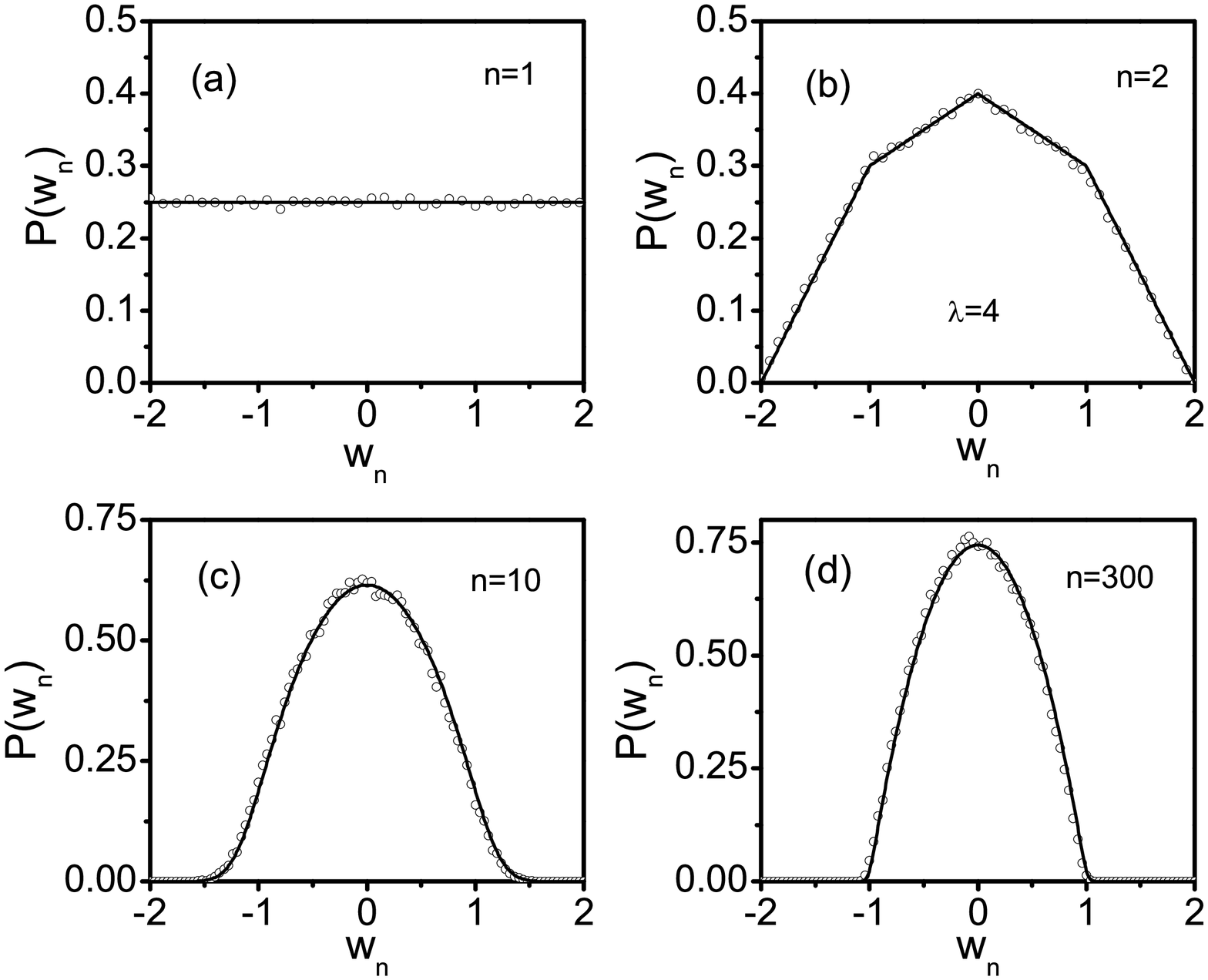}
\caption{Probability density $P(w_{n})$ of the variable $W_{n}=(1/n)%
\sum_{i=1}^{n}X_{i},$ where each random variable $X_{i}$ follows from the
composed P\'{o}lya urn scheme defined by Eqs. (\protect\ref{composed}) and (%
\protect\ref{urn1}), with $\protect\lambda =4.$ The weights [Eq. (\protect
\ref{CF2})] are $q_{+}=q_{-}=1/2.$ The solid line follows from the inverse
Fourier transform of Eq. (\protect\ref{FourierSolution}) defined with Eq. (%
\protect\ref{FourierPrimera}). The circles correspond to numerical results
obtained by averaging $5\times 10^{5}$ realizations. }
\end{figure}

The previous two expressions give an exact analytical expression for $%
G_{w}^{(n)}(k).$ For the example defined by Eq. (\ref{urn1}), the random
variables have a uniform distribution for $X\in \lbrack -2,2].$ The
probabilities of each subdomain are defined by Eq. (\ref{urn2}). Their
Fourier transform read 
\begin{equation}
G_{\pm }(k)=[\sin (k)/k]e^{\pm ik}.  \label{FourierPrimera}
\end{equation}%
In order to check these results, in Fig. 6 we show a set of probability
distributions obtained by averaging the realizations of the composed scheme
(Fig. 3). For each $n=1,$ $2,$ $10,$ $300,$ the numerical results follows
after averaging $5\times 10^{5}$ realizations. For $n=1$ it is recovered Eq.
(\ref{urn1}). For higher $n$ we find that the (numerical) inverse Fourier
transform of Eq. (\ref{FourierSolution}) evaluated with Eq. (\ref%
{FourierPrimera}) fits very well the numerical results (circles).
Consistently with the previous analysis, at $n=300$ the density $P(w_{n})$
is almost indistinguishable from the corresponding attractor, that is, $%
P(w_{n})$ in Fig. 6(d) is very well fitted by the density $P(w)$
corresponding to the standard scheme, Eq. (\ref{SpinPw}), which in turn is
plotted in Fig. 4(b). This follows because the average values $\{\bar{x}%
_{\mu }\}$ [Eq. (\ref{AveragesComposed})] are $\bar{x}_{\pm }=\pm 1$ and
also the weights $\{q_{\mu }\}$ [Eq. (\ref{CF2})]\ are $q_{\pm }=1/2,$ which
correspond to the parameters of Fig. 4. We also checked that for all values
of $\lambda $ the attractors correspond to those shown in that figure.

\begin{figure}[tbph]
\includegraphics[bb=55 560 715 1110,angle=0,width=9cm]{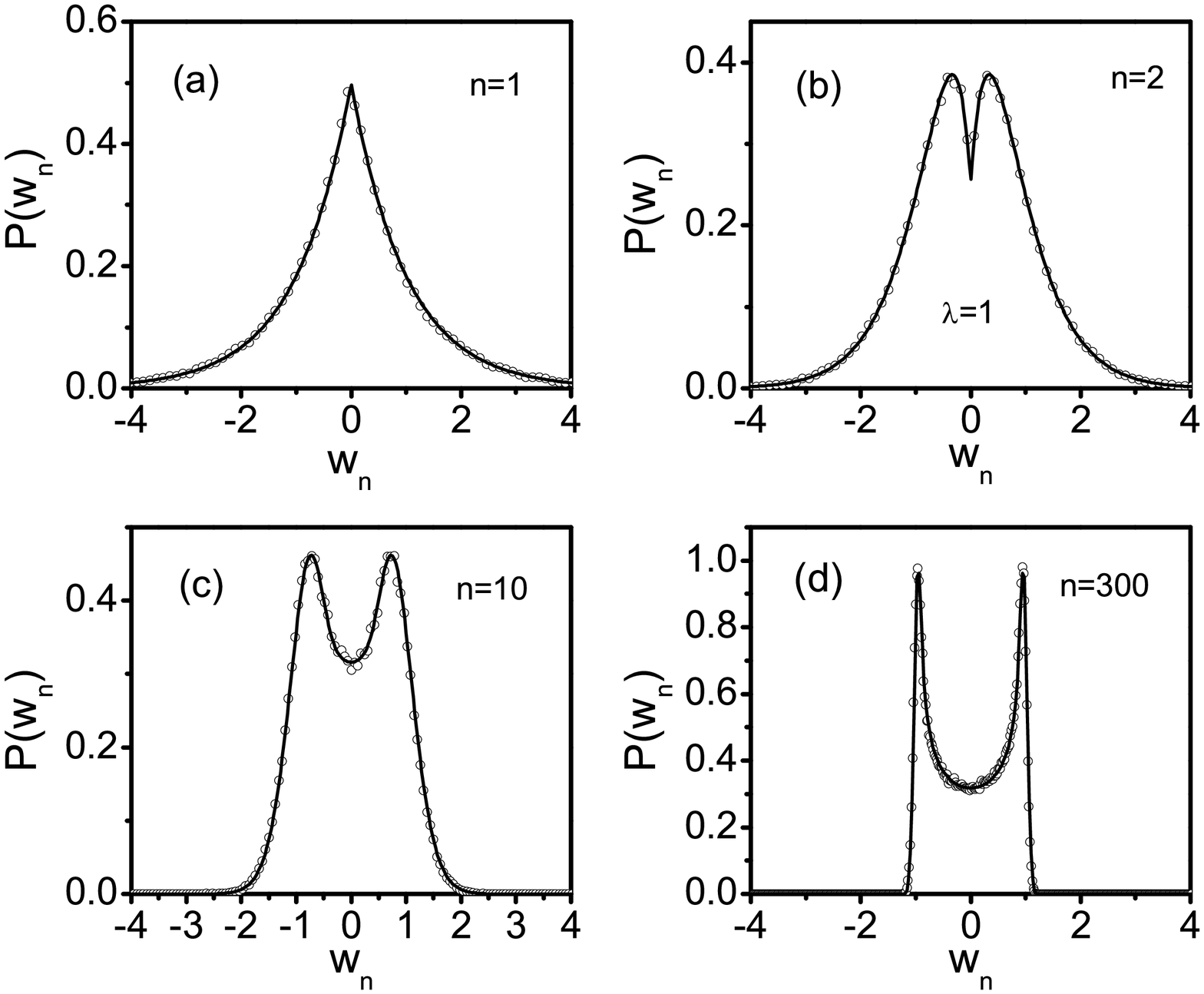}
\caption{Probability density $P(w_{n})$ of the variable $W_{n}=(1/n)%
\sum_{i=1}^{n}X_{i},$ where each random variable $X_{i}$ follows from the
composed P\'{o}lya urn scheme defined by Eqs. (\protect\ref{composed}) and (%
\protect\ref{expor}), with $\protect\lambda =1.$ The weights [Eq. (\protect
\ref{CF2})] are $q_{+}=q_{-}=1/2.$ The solid line follows from the inverse
Fourier transform of Eq. (\protect\ref{FourierSolution}) defined with Eq. (%
\protect\ref{FourierExpor}). The circles correspond to numerical results
obtained by averaging $5\times 10^{5}$ realizations. }
\end{figure}

For arbitrary distributions $P_{1}(x)$ the sum variable, associated to the
composed urn scheme with two subdomains, is characterized by the attractor
Eq. (\ref{SpinPw}). In general, the random variables can assume values over
the entire real line. For example, we take%
\begin{equation}
P_{1}(x)=(1/2)\exp (-|x|),  \label{expor}
\end{equation}%
with subdomains $\Omega _{\pm }=\{x\lessgtr 0\}.$ Then, the Fourier
transforms of $p_{\pm }(x)=\exp (\mp x)$ read%
\begin{equation}
G_{\pm }(k)=\frac{1}{1\mp ik}.  \label{FourierExpor}
\end{equation}%
In Fig. 7 we show a set of probability distributions for the sum variable
for this alternative single statistics. As in the previous case, the
analytical expressions in the Fourier domain fit the numerical results.
Notice that even when the single variables assume values over the real line,
their normalized sum is characterized by an (probability density) attractor
that is not null only in the interval $(-1,1)$ [see Fig. 4(d)]. This
property is induced by the global correlation effects.

For a urn model with three states similar results can be obtained. For
example, by maintaining $P_{1}(x)$ given by Eq. (\ref{urn1}), taking the
subdomains $\Omega _{+}=\!\{x\in (1/3,5/3)\},$ $\Omega _{-}=\!\{x\in
(-5/3,-1/3)\},$ and $\Omega _{0}=\!\{x\in (-2,-5/3)\cup (5/3,2)\}$ lead to
the attractors shown in Fig. 5. A model with exponential distributed
variables can also be written.

\section{Summary and Conclusions}

Beyond statistically independent variables, there exist very few
generalizations of the CLT. Here, we studied this problem for globally
correlated random variables that are similar and interchangeable. In order
to characterize these symmetries we derived a hierarchical set of equations
that the transition probability densities must to satisfy, Eq. (\ref%
{RecursivaCondicional}). These integral equations provide a tool for
constructing correlation mechanisms that satisfy the required properties.

Different correlations mechanisms lead to transitions probability densities
that fulfill the demanded symmetries, such as globally correlated Gaussian
variables, de Finetti representation (see Appendixes) and urn schemes. In
this last context, we introduced a generalization of P\'{o}lya urn scheme,
where the values assumed by the random variables are split in different
subdomains over the real line, each one being endowed with a probability
density. Each subdomain is chosen in agreement with the standard P\'{o}lya
urn scheme, while the associated probability density delivers the next
random value (Fig. 3). The transition probability of this composed scheme,
Eq. (\ref{composed}), fulfill the required symmetries.

The sum variable that define the CLT, Eq. (\ref{Wdefinition}), here is
defined with a different normalization because the studied random variables
may achieve a fully correlated regime. Thus, the case of independent
variables leads to a delta Dirac distribution, fact related with the
validity of the law of large numbers. In general, global correlations
consistent with the demanded symmetries lead to very different statistics of
the sum variable. The P\'{o}lya urn scheme, depending on its number of
states and characteristic parameters, delivers different probability
densities, Eq. (\ref{PwPolya}) (see Figs. 4 and 5). For two states, the
attractor is defined by an asymmetric $q$-Gaussian density $(q<1),$ Eq. (\ref%
{SpinPw}). More complex expressions arise for more states.

Given the diversity of possible attractors, a very difficult task is to
define their basin of attraction. We solved this problem in a constructive
way. We demonstrated that sum of random variables generated via the composed
P\'{o}lya urn scheme are in the basin of attraction of the distributions
associated to the standard P\'{o}lya urn scheme (see Figs. 6 and 7). This
basin is as wide as in the standard CLT. In fact, there exist infinite
single probability distributions that with a specific splitting of their
domain lead to the same attractor [see Eqs. (\ref{CF1}) and (\ref{CF2})].
The mechanism that guarantees this result is the validity of the law of
large numbers for the probability density of each subdomain as well as the
convergence to stationary values of the transition probability of the
standard P\'{o}lya urn scheme.

The basin of attraction of the P\'{o}lya urn attractors can be extended
after raising up the interchangeability symmetry in the composed scheme
[Eqs. (\ref{C1}) and (\ref{C2})]. In addition, the same attractors arise,
for example, by introducing correlations between the random variables in
such a way that the law of large numbers remains valid in each subdomain. On
the other hand, the present results lead us to ask about physical systems
characterized by dynamical correlations able to induce attractors that take
values on a subdomain of the underlying random process (variables).

In conclusions, we developed a consistent approach for dealing with globally
correlated similar interchangeable random variables, which in turn allowed
us to characterize different attractors of the CLT as well as their basin of
attraction.

\section*{Acknowledgments}

This work was supported by Consejo Nacional de Investigaciones Cient\'{\i}%
ficas y T\'{e}cnicas (CONICET), Argentina.

\appendix

\section{\label{jerarquia}Interchangeability condition for the conditional
probabilities}

Here, we derive the hierarchical set of conditions defined by Eq. (\ref%
{RecursivaCondicional}). Assuming that interchangeability is valid for $%
P_{n}(x_{1},\cdots x_{n}),$ we determine the conditions under which $%
P_{n+1}(x_{1},\cdots x_{n+1})$ also fulfill the symmetry. These functions
are related as $P_{n+1}(x_{1},\cdots x_{n+1})=P_{n}(x_{1},\cdots
x_{n})T_{n}(x_{1},\cdots x_{n}|x_{n+1}).$ Therefore, $T_{n}(x_{1},\cdots
x_{n}|x_{n+1})$ must also be symmetric in the $x_{1},\cdots x_{n}$
arguments. The interchangeability for $P_{n+1}(x_{1},\cdots x_{n+1})$ is
valid when $x_{n+1}$ can be interchanged with an arbitrary $x_{k},$ with $%
k=1,\cdots n.$ Written in an explicit way, this requirement reads%
\begin{equation}
P_{n+1}(x_{1},\cdots ,x_{k},\cdots x_{n+1})=P_{n+1}(x_{1},\cdots
,x_{n+1},\cdots x_{k}).  \label{igual}
\end{equation}%
By using Bayes rule, these objects can be written as%
\begin{equation*}
\begin{array}{r}
P_{n+1}(x_{1},\cdots ,x_{k},\cdots x_{n+1})\!=\!P_{k-1}(x_{1}\cdots x_{k-1})
\\ 
\times T_{k-1}(x_{1}\cdots x_{k-1}|x_{k}) \\ 
\times T_{k}(x_{1},\cdots x_{k}|x_{k+1}) \\ 
\times T_{k+1}(x_{1},\cdots ,x_{k},x_{k+1}|x_{k+2}) \\ 
\cdots \times T_{n-1}(x_{1},\cdots x_{n-1}|x_{n}) \\ 
\times T_{n}(x_{1},\cdots x_{n}|x_{n+1}),%
\end{array}%
\end{equation*}%
and also%
\begin{equation*}
\begin{array}{r}
P_{n+1}(x_{1},\cdots ,x_{n+1},\cdots x_{k})\!=\!P_{k-1}(x_{1}\cdots x_{k-1})
\\ 
\cdots \times T_{k-1}(x_{1}\cdots x_{k-1}|x_{n+1}) \\ 
\times T_{k}(x_{1},\cdots ,x_{k-1},x_{n+1}|x_{k+1}) \\ 
\times T_{k+1}(x_{1},\cdots ,x_{k-1},x_{n+1},x_{k+1}|x_{k+2}) \\ 
\cdots \times T_{n-1}(x_{1},\cdots ,x_{k-1},x_{n+1},x_{k+1}\cdots
x_{n-1}|x_{n}) \\ 
T_{n}(x_{1},\cdots ,x_{k-1},x_{n+1},x_{k+1}\cdots x_{n-1}|x_{k}),%
\end{array}%
\end{equation*}%
where now $k=2,\cdots n.$ Performing the integrals $\int
dx_{k}dx_{k+1}\cdots dx_{n}$ to both objects, using the normalization
condition $\int dx_{j}T_{i}(x_{1},\cdots x_{i}|x_{j})=1,$ and simplifying
the factor $P_{k-1}(x_{1}\cdots x_{k-1}),$ from Eq. (\ref{igual}) it follows
the condition%
\begin{equation}
\begin{array}{r}
T_{k-1}(x_{1}\cdots x_{k-1}|x_{n+1})=\int dx_{k}\cdots dx_{n} \\ 
T_{k-1}(x_{1}\cdots x_{k-1}|x_{k}) \\ 
\times T_{k}(x_{1},\cdots x_{k}|x_{k+1}) \\ 
\times T_{k+1}(x_{1},\cdots ,x_{k},x_{k+1}|x_{k+2}) \\ 
\cdots \times T_{n-1}(x_{1},\cdots x_{n-1}|x_{n}) \\ 
\times T_{n}(x_{1},\cdots x_{n}|x_{n+1}).%
\end{array}
\label{muchas}
\end{equation}%
For $k=n,$ this equation reduces to%
\begin{eqnarray}
T_{n-1}(x_{1}\cdots x_{n-1}|x_{n+1}) &=&\int dx_{n}T_{n-1}(x_{1}\cdots
x_{n-1}|x_{n})  \notag \\
&&\times T_{n}(x_{1},\cdots x_{n}|x_{n+1}).  \label{TN}
\end{eqnarray}%
For $k=n-1,$ after using the validity of Eq. (\ref{TN}), Eq. (\ref{muchas})
leads to%
\begin{eqnarray*}
T_{n-2}(x_{1}\cdots x_{n-2}|x_{n+1})\! &=&\!\int dx_{n-1} \\
&&T_{n-2}(x_{1}\cdots x_{n-2}|x_{n-1}) \\
&&\times T_{n-1}(x_{1},\cdots x_{n-1}|x_{n+1}).
\end{eqnarray*}%
Notice that this equation has the same structure as Eq. (\ref{TN}). Hence,
it is simple to realize that Eq. (\ref{muchas}) is satisfied if%
\begin{eqnarray*}
\!T_{n-j}(x_{1}\cdots x_{n-j}|x_{n+1})\! &=&\!\int dx_{n-j+1} \\
&&T_{n-j}(x_{1}\cdots x_{n-j}|x_{n-j+1}) \\
&&\times T_{n-j+1}(x_{1},\cdots x_{n-j+1}|x_{n+1}).\ \ \ \ \ \ 
\end{eqnarray*}%
where $j=1,\cdots n-(k-1).$ This last equation, after a straightforward
change of indexes, recovers Eq. (\ref{RecursivaCondicional}).

\section{\label{additive}Additive memory case}

The symmetry of the transition probability $T_{n}(x_{1}\cdots x_{n}|x_{n+1})$
on the previous conditional values $x_{1}\cdots x_{n}$ is trivially
fulfilled by assuming that it depends on the addition of these values. Then,
we write%
\begin{equation}
T_{n}(x_{1}\cdots x_{n}|x_{n+1})=\mathcal{T}_{n}(x_{1}+x_{2}\cdots
+x_{n}|x_{n+1}),
\end{equation}%
where $\mathcal{T}_{n}(x^{\prime }|x)$ is a set of equivalent functions that
only depends on two arguments. For random variables $\{X_{i}\}_{i=1}^{n}$
with a finite support, $X\in \lbrack x_{<},x_{>}],$ the variable $x^{\prime
} $ in $\mathcal{T}_{n}(x^{\prime }|x)$ runs in the interval $%
[nx_{<},nx_{>}].$

From Eq. (\ref{RecursivaCondicional}), it follows that the functions $%
\mathcal{T}_{n}(x^{\prime }|x)$ must to satisfy the recursive relations 
\begin{equation}
\mathcal{T}_{n-1}(x^{\prime }|x)=\int dy\mathcal{T}_{n-1}(x^{\prime }|y)%
\mathcal{T}_{n}(x^{\prime }+y|x).  \label{CondicionalLineal}
\end{equation}%
Below we show that the additive assumption allows us to find a complete
solution of the hierarchy (\ref{RecursivaCondicional}) after assuming
different statistics for each single variable.

\subsection{Gaussian random variables}

For the single distribution of each random variable, let assume a Gaussian
distribution%
\begin{equation}
P_{1}(x)=\frac{1}{\sqrt{2\pi \sigma ^{2}}}\exp \Big{[}-\frac{x^{2}}{2\sigma
^{2}}\Big{]},  \label{P1Gauss}
\end{equation}%
which satisfies $\int dxP_{1}(x)=1.$ The width $\sigma ^{2}$ is a free
parameter. Given that $\mathcal{T}_{1}(x^{\prime }|x)=T_{1}(x^{\prime }|x),$
in order to fulfill Eq. (\ref{CondicionalOne}) we assume that $\mathcal{T}%
_{1}(x^{\prime }|x)$\ is a Gaussian distribution in both variables $%
x^{\prime }$ and $x.$ Hence, $\mathcal{T}_{1}(x^{\prime }|x)\approx \exp [-(%
\frac{x^{2}}{2\rho ^{2}}+\frac{x^{\prime 2}}{2\mu ^{2}}+\frac{xx^{\prime }}{%
\nu })].$ The undetermined free parameters $(\rho ,\mu ,\nu )$ are
constrained by the normalization condition $\int dx\mathcal{T}_{1}(x^{\prime
}|x)=1,$ and Eq. (\ref{CondicionalOne}). After imposing these constraints,
we obtain%
\begin{equation}
\mathcal{T}_{1}(x^{\prime }|x)=\frac{1}{\sqrt{2\pi \rho ^{2}}}\exp \Big{[}-%
\frac{1}{2\rho ^{2}}(x-\lambda x^{\prime })^{2}\Big{]},  \label{T1Gauss}
\end{equation}%
where the real parameter $\lambda $ is%
\begin{equation}
\lambda \equiv \sqrt{1-\frac{\rho ^{2}}{\sigma ^{2}}.}  \label{lambda}
\end{equation}%
$\rho $ remains as a free parameter and satisfies $\rho ^{2}\leq \sigma
^{2}. $ Notice that when $\lambda =0,$ that is $\rho =\sigma ,$ we get
independent variables, $\mathcal{T}_{1}(x^{\prime }|x)=P_{1}(x).$ On the
other hand, for $\lambda =1,$ $\rho \rightarrow 0,$ it follows $\mathcal{T}%
_{1}(x^{\prime }|x)=\delta (x-x^{\prime }).$ This is the maximal correlated
case, where $x=x^{\prime }.$ Hence, after the first random value, the next
one is equal to the previous one.

Higher transition probabilities can be obtained from Eq. (\ref%
{CondicionalLineal}) and the solution (\ref{T1Gauss}). Proposing a Gaussian
structure for higher objects, we get%
\begin{equation}
\mathcal{T}_{n}(x^{\prime }|x)=\frac{1}{\sqrt{2\pi \rho _{n}^{2}}}\exp %
\Big{[}-\frac{1}{2\rho _{n}^{2}}(x-\lambda _{n}x^{\prime })^{2}\Big{]},
\label{ConditionalGauss}
\end{equation}%
where the coefficients satisfy the recursive relations 
\begin{equation}
\lambda _{n}=\frac{\lambda _{n-1}}{1+\lambda _{n-1}},\ \ \ \ \ \ \ \ \ \rho
_{n}^{2}=\Big{[}1-\Big{(}\frac{\lambda _{n-1}}{1+\lambda _{n-1}}\Big{)}^{2}%
\Big{]}\rho _{n-1}^{2},  \label{Recursivas}
\end{equation}%
$(n\geq 2),$ with $\lambda _{1}\equiv \lambda $ and $\rho _{1}\equiv \rho .$
Their solution is%
\begin{equation}
\lambda _{n}=\frac{\lambda }{1+(n-1)\lambda },\ \ \ \ \ \ \rho _{n}^{2}=%
\Big{[}1+\frac{(n-1)\lambda ^{2}}{1+n\lambda }\Big{]}^{-1}\rho ^{2}.
\label{LambdaRho}
\end{equation}

The joint probability distribution $P_{n}(x_{1},\cdots x_{n})$ can be
obtained from the set of transition probabilities [Eq. (\ref{Conjunta})].
For example, the joint probability $P_{2}(x_{1},x_{2}),$ from Eqs. (\ref%
{P1Gauss}) and (\ref{T1Gauss}), reads%
\begin{equation}
P_{2}(x_{1},x_{2})=\frac{1}{2\pi \sqrt{\sigma ^{2}\rho ^{2}}}\exp \Big{[}-%
\frac{1}{2\rho ^{2}}(x_{1}^{2}+x_{2}^{2}-2\lambda x_{1}x_{2})\Big{]},
\end{equation}%
which consistently is symmetric in $x_{1}$ and $x_{2}.$ For arbitrary $n\geq
2,$ we get%
\begin{equation}
P_{n}(x_{1},\cdots x_{n})=\sqrt{\frac{\det [A^{(n)}]}{(2\pi )^{n}}}\exp %
\Big{[}-\frac{1}{2}\sum_{i,j=1}^{n}x_{i}\mathrm{A}_{ij}^{(n)}x_{j}\Big{]},
\label{JointGauss}
\end{equation}%
where the matrix elements are%
\begin{equation}
\mathrm{A}_{ii}^{(n)}=\frac{1}{\rho _{n-1}^{2}},\ \ \ \ \ \ \ \mathrm{A}%
_{ij}^{(n)}=-\frac{\lambda _{n-1}}{\rho _{n-1}^{2}},\ \ \ \ i\neq j,
\label{MatrizA}
\end{equation}%
where $\rho _{n}$ and $\lambda _{n}$ are defined by Eq. (\ref{LambdaRho}).
The determinant of the matrix $\mathrm{A}_{ij}^{(n)}$ reads%
\begin{equation}
\det [\mathrm{A}^{(n)}]=\Big{\{}\lbrack 1+(n-1)\lambda ]\sigma ^{2}\Big{(}%
\frac{\rho ^{2}}{1+\lambda }\Big{)}^{n-1}\Big{\}}^{-1}.
\end{equation}%
The validity of Eq. (\ref{JointGauss}) can be probe by using the
mathematical principle of induction and the recursive relations (\ref%
{Recursivas}).

We remark that Eq. (\ref{JointGauss}) was derived over the basis of the
conditional probabilities densities (\ref{ConditionalGauss}), which in turn
are a solution of the hierarchy (\ref{CondicionalLineal}) after assuming the
Gaussian statistics defined by Eq. (\ref{P1Gauss}). Clearly, due to the
symmetry of the covariance matrix (\ref{MatrizA}), the multidimensional
Gaussian density (\ref{JointGauss}) is compatible with the
interchangeability symmetry.

Now we obtain the distribution of $W$ [Eq. (\ref{Wdefinition})]\ for a set
of random variables $\{X_{i}\}_{i=1}^{n}$ correlated in agreement with the
Gaussian distribution Eq. (\ref{JointGauss}), which in turn is related to
the transition probability Eq. (\ref{ConditionalGauss}). The (multiple)
Fourier transform of Eq. (\ref{JointGauss}) reads%
\begin{equation}
G_{k}(k_{1},\cdots k_{k})=\exp \Big{[}-\frac{1}{2}\sum_{i,j=1}^{k}k_{i}(1/%
\mathrm{A}^{(k)})_{ij}k_{j}\Big{]},  \label{GaussConjuntaFourier}
\end{equation}%
where $(1/\mathrm{A}^{(k)})$ is the matrix inverse of $\mathrm{A}^{(k)}$
[Eq. (\ref{MatrizA})]. It can be written as%
\begin{equation}
(1/\mathrm{A}^{(k)})_{ii}=\sigma ^{2},\ \ \ \ \ \ \ (1/\mathrm{A}%
^{(k)})_{ij}=\sigma ^{2}\lambda ,\ \ \ \ i\neq j,
\end{equation}%
where $\lambda =(1-\rho ^{2}/\sigma ^{2})^{1/2}$ [Eq. (\ref{lambda})].
Hence, from Eqs. (\ref{FourierSuma}) and (\ref{GaussConjuntaFourier}), we get%
\begin{equation}
G_{w_{n}}(k)=\exp \Big{\{}-\frac{1}{2}\sigma ^{2}\lambda k^{2}\Big{[}1+\frac{%
1}{n}(\lambda ^{-1}-1)\Big{]}\Big{\}}.
\end{equation}%
After taking the limit $n\rightarrow \infty ,$ it follows%
\begin{equation}
P(w)=\sqrt{\frac{1}{2\pi \sigma ^{2}\lambda }}\exp \Big{[}-\frac{1}{2}\frac{%
w^{2}}{\sigma ^{2}\lambda }\Big{]}.  \label{PGaussol}
\end{equation}%
Contrarily to the case of independent variables, here the distribution of $W$
is not a delta Dirac distribution, Eq. (\ref{DeltaIndependientes}). This
departure has its origin in the correlations between the random variables,
which are tuned by the parameter $\lambda .$ In fact, in the limit $\lambda
\rightarrow 0$ we recover Eq. (\ref{DeltaIndependientes}) with $\bar{x}=0,$
that is, independent variables. On the other hand, for maximally correlated
variables, $\lambda \rightarrow 1,$ we recover the Gaussian distribution $%
P_{1}(x)$ [Eq. (\ref{P1Gauss})]. This result, which gives the maximal
departure with respect to independent variables, follows after noting that
all random variables assume the same value [see the transition probabilities
Eqs. (\ref{T1Gauss}) and (\ref{ConditionalGauss})].

\subsection{Linear additive memory case}

Here, we search another class of solution which in addition assume that the
transition probabilities\ $\mathcal{T}_{n}(x^{\prime }|x)$ depend linearly
on the argument $x^{\prime }.$ In the following results, the structure of $%
P_{1}(x)$ is arbitrary.

Given $P_{1}(x),$ and given the linear dependence of $\mathcal{T}%
_{1}(x^{\prime }|x)$ on $x^{\prime },$ the relation defined by Eq. (\ref%
{CondicionalOne}) becomes%
\begin{equation}
\mathcal{T}_{1}(\left\langle X\right\rangle |x)=P_{1}(x),\ \ \ \ \ \ \ \
\left\langle X\right\rangle \equiv \int dxP_{1}(x)x.  \label{Tekila}
\end{equation}%
Given $P_{1}(x),$ any transition probability density $T_{1}(x^{\prime }|x)$\
satisfying this equation is a valid one. On the other hand, assuming that
all transition probability densities depend linearly on $x^{\prime },$ the
conditions (\ref{CondicionalLineal}) can be written as%
\begin{equation}
\mathcal{T}_{n}(x^{\prime }+\left\langle X\right\rangle _{n-1,x^{\prime
}}|x)=\mathcal{T}_{n-1}(x^{\prime }|x),  \label{RecursivaLinealCase}
\end{equation}%
where the conditional\ average $\left\langle X\right\rangle _{n-1,x^{\prime
}}$ is defined as%
\begin{equation}
\left\langle X\right\rangle _{n-1,x^{\prime }}\equiv \int dx\mathcal{T}%
_{n-1}(x^{\prime }|x)x.  \label{ConditionalAverage}
\end{equation}%
By evaluating the previous two expressions in $x^{\prime }=\left\langle
X\right\rangle ,$ it follows the relation%
\begin{equation}
\mathcal{T}_{n}(n\left\langle X\right\rangle |x)=P_{1}(x),
\end{equation}%
which generalize that defined by Eq. (\ref{Tekila}).

From Eq. (\ref{ConditionalAverage}), we realize that $\left\langle
X\right\rangle _{n-1,x^{\prime }}$ is also a linear function of $x^{\prime
}. $ In particular, it is possible to write%
\begin{equation}
\left\langle X\right\rangle _{1,x^{\prime }}=\int dx\mathcal{T}%
_{1}(x^{\prime }|x)x=ax^{\prime }+b.  \label{LinearAverage}
\end{equation}%
This equation defines the constants $a$ and $b,$ the former being a
dimensionless one, while the last one has units of $x.$ Multiplying the
previous expression by $P_{1}(x^{\prime })$ and integrating in $x^{\prime }$
it follows the relation $\left\langle X\right\rangle =b/(1-a).$

From (\ref{LinearAverage}), the solution of Eq. (\ref{RecursivaLinealCase})
for $n=2$ is $\mathcal{T}_{2}(x^{\prime }(1+a)+b|x)=\mathcal{T}%
_{1}(x^{\prime }|x),$\ which can be rewritten as%
\begin{equation}
\mathcal{T}_{2}(x^{\prime }|x)=\mathcal{T}_{1}\Big{(}\frac{x^{\prime }-b}{1+a%
}\Big{|}x\Big{)}.
\end{equation}%
In a similar form, an explicit expression for $\mathcal{T}_{3}(x^{\prime
}|x) $ can be obtained. For arbitrary $n,$ as a solution of Eq. (\ref%
{RecursivaLinealCase}) we propose the expression%
\begin{equation}
\mathcal{T}_{n}(x^{\prime }|x)=\mathcal{T}_{1}\Big{(}\frac{x^{\prime }-(n-1)b%
}{1+(n-1)a}\Big{|}x\Big{)}.  \label{Solutor}
\end{equation}%
The validity of this result can be prove from Eq. (\ref{RecursivaLinealCase}%
) by using the mathematical principle of induction.

\subsubsection*{Discrete distributions with finite support}

The set of functions defined by Eq. (\ref{Solutor}) give a full solution to
the hierarchical structure (\ref{CondicionalLineal}). Nevertheless, it is
not guaranteed that their are positive functions. In order to check this
issue, we consider discrete random variables defined by 
\begin{equation}
P_{1}(x)=\sum\nolimits_{\mu =1}^{M}q_{\mu }\delta (x-x_{\mu }),
\end{equation}%
where the positive weights\ satisfy $\sum\nolimits_{\mu =1}^{M}q_{\mu }=1.$

The mean value, $\left\langle X\right\rangle =\int dxP_{1}(x)x,$ reads $%
\left\langle X\right\rangle =\sum\nolimits_{\mu =1}^{M}q_{\mu }x_{\mu }.$
The first conditional density, given it linear dependence on $x^{\prime },$
is written as%
\begin{equation}
\mathcal{T}_{1}(x^{\prime }|x)=\frac{1}{\mathcal{N}}\sum\nolimits_{\mu
=1}^{M}(\alpha _{\mu }+\beta _{\mu }x^{\prime })\delta (x-x_{\mu }),
\end{equation}%
where $(\alpha _{\mu },\beta _{\mu })$ and $\mathcal{N}$ are arbitrary
parameters. Using the normalization condition $\int dx\mathcal{T}%
_{1}(x^{\prime }|x)=1,$ it follows $\mathcal{N}=\sum\nolimits_{\mu
=1}^{M}\alpha _{\mu },$ and%
\begin{equation}
\sum\nolimits_{\mu =1}^{M}\beta _{\mu }=0.
\end{equation}%
The condition $\mathcal{T}_{1}(\left\langle X\right\rangle |x)=P_{1}(x),$
leads to $\frac{1}{\mathcal{N}}(\alpha _{\mu }+\beta _{\mu }\left\langle
X\right\rangle )=q_{\mu }.$ Under the association $(\beta _{\mu }/\mathcal{N}%
)\rightarrow \beta _{\mu },$ we get%
\begin{equation}
\mathcal{T}_{1}(x^{\prime }|x)=\sum\nolimits_{\mu =1}^{M}[q_{\mu }+\beta
_{\mu }(x^{\prime }-\left\langle X\right\rangle )]\delta (x-x_{\mu }).
\label{T1Suma}
\end{equation}%
The first conditional average reads%
\begin{equation}
\int dx\mathcal{T}_{1}(x^{\prime }|x)x=\zeta x^{\prime }+\left\langle
X\right\rangle (1-\zeta )=ax^{\prime }+b,  \label{ayb}
\end{equation}%
where the constant $\zeta $\ is%
\begin{equation}
\zeta \equiv \sum\nolimits_{\mu =1}^{M}x_{\mu }\beta _{\mu }.
\end{equation}%
From Eq. (\ref{Solutor}), higher objects reads%
\begin{equation}
\mathcal{T}_{n}(x^{\prime }|x)=\sum\nolimits_{\mu =1}^{M}\Big{[}q_{\mu
}+\beta _{\mu }\frac{x^{\prime }-n\left\langle X\right\rangle }{1+(n-1)\zeta 
}\Big{]}\delta (x-x_{\mu }).  \label{Solutora}
\end{equation}%
We remark that this set of equations provide a solution to the full
hierarchy of conditional probabilities under the interchangeability
symmetry. Nevertheless, the positivity of these objects must to be checked.

The constants $\beta _{\mu }$ should be chosen such that the positivity of $%
\mathcal{T}_{n}(x^{\prime }|x)$ is guaranteed for all $n$ and $x^{\prime
}\in (nx_{<},nx_{>}),$ where $x_{<}$ and $x_{>}$ define respectively the
minimal and maximal values of the set $\{x_{\mu }\}_{\mu =1}^{M}.$ Hence,
for $n=1$ it follows%
\begin{equation}
q_{\mu }+\beta _{\mu }(x-\left\langle X\right\rangle )\geq 0,  \label{Uno}
\end{equation}%
while in the limit $n\rightarrow \infty ,$ we get%
\begin{equation}
q_{\mu }+\frac{\beta _{\mu }(x-\left\langle X\right\rangle )}{%
\sum\nolimits_{\nu =1}^{M}x_{\nu }\beta _{\nu }}\geq 0.  \label{Infty}
\end{equation}%
In both inequalities, $x$ assume values over the set $\{x_{u}\}.$ In the
case of two states, $M=2,$ from these inequalities we obtain $\beta \leq
|x_{2}-x_{1}|^{-1},$ where $\beta _{1}=-\beta _{2}=\beta ,$ and $\{x_{\mu
}\}=\{x_{1},x_{2}\}.$ Hence, positivity can always be guaranteed in this
case.

In general for $M\geq 3,$ there is not a solution for the set $\{\beta _{\mu
}\}$ that guarantees the validity of the previous two inequalities. In fact,
from Eq. (\ref{Infty}), we deduce that%
\begin{equation}
|\beta _{\mu }(x-\left\langle X\right\rangle )|\leq q_{\mu }\left\vert
\sum\nolimits_{\nu =1}^{M}x_{\nu }\beta _{\nu }\right\vert .
\end{equation}%
Taking $x\rightarrow x_{\mu },$ and adding in the $\mu $ index, $%
\sum\nolimits_{\mu =1}^{M},$ it follows%
\begin{equation}
\sum\nolimits_{\nu =1}^{M}|\beta _{\mu }x_{\mu }|\leq \left\vert
\sum\nolimits_{\nu =1}^{M}x_{\mu }\beta _{\mu }\right\vert .
\end{equation}%
Hence, we deduce that $x_{\mu }\beta _{\mu }>0,$ and then $%
\sum\nolimits_{\mu =1}^{M}x_{\mu }\beta _{\mu }>0.$ Therefore, Eqs. (\ref%
{Uno}) and (\ref{Infty}) are equivalents, in the sense that one of them
always implies the other. Taking one of them and the previous one, it
follows $M(M-1)-1$ equations, while the number of variables is $M-1.$ Thus,
a consistent solution (positive transition probabilities) is \textit{only}
available when $M=2.$

For classical spin variables $x_{\mu }=\pm 1,$ parametrizing $\beta
=(1/2)(1+\lambda )^{-1}\leq |x_{+}-x_{-}|^{-1}=1/2,$ from Eq. (\ref{Solutora}%
) we get $(\lambda _{\pm }=\lambda q_{\pm })$%
\begin{equation}
\mathcal{T}_{n}(x^{\prime }|x)\!=\!\Big{(}\!\frac{\lambda _{+}+\frac{%
n+x^{\prime }}{2}}{n+\lambda }\!\Big{)}\delta (x-1)+\Big{(}\!\frac{\lambda
_{-}+\frac{n-x^{\prime }}{2}}{n+\lambda }\!\Big{)}\delta (x+1).
\label{TN_TLS}
\end{equation}%
This expression gives a positive solution consistent with
interchangeability. Nevertheless, it is simple to realize that the
quantities $\frac{n+x^{\prime }}{2}$ and $\frac{n-x^{\prime }}{2}$ give the
number of times $n_{+}$ and $n_{-}$ that the previous variables assumed the
values $\pm 1$ respectively. Therefore, Eq. (\ref{TN_TLS}) recovers the
transition probability corresponding to the P\'{o}lya urn scheme, Eq. (\ref%
{TransitionDirichlet}).

\section{\label{deFinetti}de Finetti representation}

de Finetti \cite{finetti} introduced the concept of interchangeability and
also defined a general representation structure for the joint probability
density of a set of dichotomic interchangeable variables. The de Finetti
representation can be generalized for arbitrary (non-dichotomic) random
variables. Given a set of interchangeable random variables $%
\{X_{i}\}_{i=1}^{n},$ their $n$-joint probability density is expressed as%
\begin{equation}
P_{n}(x_{1},\cdots ,x_{n})=\int_{\Omega _{y}}dyp(y)\prod_{i=1}^{n}p(y|x_{i}).
\label{FinettiGral}
\end{equation}%
Here, $p(y)$ is the probability density of an extra random variable\ $Y,$
which assume values in the domain $\Omega _{y}.$ On the other hand, $%
p(y|x_{i})$ is a transition probability: it gives the probability density of 
$X_{i}$ given the value $y$ of the random variable $Y.$

The structure given by Eq. (\ref{FinettiGral}) allows us to read the
realizations of the correlated set $\{X_{i}\}_{i=1}^{n}$\ as an average over
realizations of a set of identical random variables with the joint
probability density $\prod_{i=1}^{n}p(y|x_{i}).$ A similar interpretation
can be obtained for the transition density $T_{n}(x_{1},\cdots
x_{n}|x_{n+1}).$ In fact, by using that $T_{n}(x_{1},\cdots
x_{n}|x_{n+1})=P_{n+1}(x_{1},\cdots ,x_{n+1})/P_{n}(x_{1},\cdots ,x_{n}),$
it can be written as%
\begin{equation}
T_{n}(x_{1},\cdots x_{n}|x_{n+1})=\int_{\Omega _{y}}dyp_{n}(\{x_{i}\}|y)\
p(y|x_{n+1}),  \label{TransitionAveraging}
\end{equation}%
where $p(y|x_{n+1})$ was introduced previously while $p_{n}(\{x_{i}\}|y)$ is%
\begin{equation}
p_{n}(\{x_{i}\}|y)=\frac{\prod_{i=1}^{n}p(y|x_{i})}{\int_{\Omega
_{y}}dy^{\prime }p(y^{\prime })\prod_{j=1}^{n}p(y^{\prime }|x_{j})}p(y).
\end{equation}%
Therefore, $T_{n}(x_{1},\cdots x_{n}|x_{n+1})$ is set by $p(y|x_{n+1}),$
where now the statistical distribution $p_{n}(\{x_{i}\}|y)$ of the random
variable $Y$ [see Eq. (\ref{TransitionAveraging})] depends on all previous
values $\{X_{i}\}_{i=1}^{n}.$ Hence, $p_{n}(x^{\prime }|y)$ can be read as
the conditional probability density of the random variable $Y$
\textquotedblleft given\textquotedblright\ the previous history defined by
the set of values $\{x_{i}\}_{i=1}^{n}.$ On the other hand, it is simple to
check that Eq. (\ref{TransitionAveraging}) satisfies the hierarchical
equations defined by Eq. (\ref{RecursivaCondicional}).

The sum variable (\ref{Wdefinition}) can be straightforwardly characterized
from Eqs. (\ref{FinettiGral}) and (\ref{GwFinal}). We get%
\begin{equation}
G_{w_{n}}(k)=\int_{\Omega _{y}}dyp(y)[G(y|k/n)]^{n},
\end{equation}%
where $G(y|k)\equiv \int_{-\infty }^{+\infty }dxe^{ikx}p(y|x).$ In the
asymptotic limit, assuming valid the law of large numbers for the transition 
$p(y|x),$ from Eqs. (\ref{GIndependiente}) and (\ref{DeltaIndependientes})
it follows%
\begin{equation}
P(w)=\int_{\Omega _{y}}dyp(y)\delta (w-\bar{x}_{y}),  \label{PSumaFinetti}
\end{equation}%
where the mean value $\bar{x}_{y}$\ is a function of $y,$%
\begin{equation}
\bar{x}_{y}\equiv \int dxp(y|x)x.  \label{XmedioCondicional}
\end{equation}

In the case of dichotomic variables, $X_{i}=0,1,$ with transition
probability $p(y|x_{i})=y^{x_{i}}(1-y)^{1-x_{i}},$ the joint probability $%
P_{n}(x_{1},\cdots ,x_{n}),$ from Eq. (\ref{FinettiGral}), becomes%
\begin{equation}
P_{n}(x_{1},\cdots
,x_{n})=\int_{0}^{1}dyp(y)\prod_{i=1}^{n}y^{x_{i}}(1-y)^{1-x_{i}}.
\label{ConjuntaFine}
\end{equation}%
Noting that the dependence of the probability $P_{n}(x_{1},\cdots ,x_{n})$
on the set $\{x_{i}\}_{i=1}^{n}$\ can be written in terms of the the
variable $x^{\prime }\equiv \sum_{i=1}^{n}x_{i}$ [Eq. (\ref{ConjuntaFine})],
from Eq. (\ref{TransitionAveraging}) it follows the presentation%
\begin{equation}
\mathcal{T}_{n}(x^{\prime }|x)=\int_{0}^{1}dyp_{n}(x^{\prime }|y)\
y^{x}(1-y)^{1-x},  \label{TransitionFinetti}
\end{equation}%
where%
\begin{equation}
p_{n}(x^{\prime }|y)=\frac{y^{x^{\prime }}(1-y)^{n-x^{\prime }}}{%
\int_{0}^{1}d\tilde{y}p(\tilde{y})\ \tilde{y}^{x^{\prime }}(1-\tilde{y}%
)^{n-x^{\prime }}}p(y).  \label{ConditionalY}
\end{equation}%
Eq. (\ref{TransitionFinetti}) provides a representation for the transition
probability $\mathcal{T}_{n}(x^{\prime }|x)$ similar to that defined by Eq. (%
\ref{ConjuntaFine}).

Given that Eq. (\ref{XmedioCondicional}) leads to $\bar{x}_{y}=y,$ from Eq. (%
\ref{PSumaFinetti}) it follows that $P(w)=p(y)|_{y=w}.$ Hence, any attractor
can be obtained by choosing an arbitrary density $p(y).$

If one choose a Beta distribution%
\begin{equation}
p(y)=\frac{\Gamma (\alpha +\alpha ^{\prime })}{\Gamma (\alpha )\Gamma
(\alpha ^{\prime })}y^{\alpha -1}(1-y)^{\alpha ^{\prime }-1},  \label{Beta}
\end{equation}%
where $\alpha >1$ and $\alpha ^{\prime }>1$\ are real parameters, from Eqs. (%
\ref{ConjuntaFine}) it is possible to obtain the joint probability
densities. In particular, $P_{1}(x)$ can be written as $P_{1}(x)=[\alpha
^{\prime }\delta (x)+\alpha \delta (x-1)]/(\alpha +\alpha ^{\prime }).$ On
the other hand, by rewriting Eq. (\ref{TransitionFinetti}) as $\mathcal{T}%
_{n}(x^{\prime }|x)=\delta (x)\int_{0}^{1}dyp_{n}(x^{\prime }|y)(1-y)+\delta
(x-1)\int_{0}^{1}dyp_{n}(x^{\prime }|y)y,$ the transition probability
density explicitly reads%
\begin{equation}
\mathcal{T}_{n}(x^{\prime }|x)=\frac{(n-x^{\prime }+\alpha ^{\prime })\delta
(x)+(x^{\prime }+\alpha )\delta (x-1)}{n+\alpha +\alpha ^{\prime }}.
\label{Tran2Finetti}
\end{equation}%
In deriving this expression we used the dichotomic property of the random
variables.

By introducing the parameter $\lambda =\alpha +\alpha ^{\prime },$ the
weights $q_{0}=\alpha ^{\prime }/(\alpha +\alpha ^{\prime }),$ $q_{1}=\alpha
/(\alpha +\alpha ^{\prime }),$ and the numbers $n_{0}=n-x^{\prime },\
n_{1}=x^{\prime },$ the transition probability (\ref{Tran2Finetti}) can be
written as a particular case of the P\'{o}lya urn scheme [see Eq. (\ref%
{TransitionDirichlet})]. In fact, $n_{0}$ and $n_{1}$ are the number of
times that the random variables assumed the values $0$ and $1$ respectively.

The probability of the variable $X^{\prime }\equiv \sum_{i=1}^{n}X_{i},$
from Eqs. (\ref{ConjuntaFine}) and (\ref{Beta}) reads%
\begin{equation}
P(x^{\prime })=\binom{n}{x^{\prime }}\frac{\Gamma (\alpha +\alpha ^{\prime })%
}{\Gamma (n+\alpha +\alpha ^{\prime })}\frac{\Gamma (n-x^{\prime }+\alpha
^{\prime })}{\Gamma (\alpha ^{\prime })}\frac{\Gamma (x^{\prime }+\alpha )}{%
\Gamma (\alpha )},  \label{ProbNmas}
\end{equation}%
where $\Gamma (x)$ is the Gamma function. The factor $\binom{n}{x^{\prime }}$
follows from all configurations that lead to the same value of $x^{\prime }.$

\section{\label{MacQueen}Blackwell-MacQueen urn scheme}

Here, we obtain the joint probability of the Blackwell-MacQueen urn scheme
[Eq. (\ref{TransitionQueen})], as well as the characteristic function of the
sum variable.

The probability density of $X_{1}$ is $P_{1}(x_{1}).$\ The second joint
probability density, from Eq. (\ref{Conjunta}), reads%
\begin{equation}
P_{2}(x_{1},x_{2})=\frac{\lambda p(x_{1})p(x_{2})+p(x_{1})\delta
(x_{2}-x_{1})}{(1+\lambda )}.
\end{equation}%
Furthermore,%
\begin{eqnarray}
P_{3}(x_{1},x_{2},x_{3}) &=&\frac{1}{(1+\lambda )(2+\lambda )}\Big{[}\lambda
^{2}p(x_{1})p(x_{2})p(x_{3})  \notag \\
&&+\lambda p(x_{1})p(x_{2})\delta (x_{3}-x_{2})  \notag \\
&&+\lambda p(x_{1})\delta (x_{2}-x_{1})p(x_{3}) \\
&&+\lambda p(x_{1})p(x_{2})\delta (x_{3}-x_{1})  \notag \\
&&+2p(x_{1})\delta (x_{2}-x_{1})\delta (x_{3}-x_{1})\Big{]}.  \notag
\end{eqnarray}%
In general, we can write%
\begin{equation}
P_{n}(x_{1},\cdots x_{n})=\sum_{\pi \in \Pi _{n}}w_{n}(\pi )\mathcal{P}_{\pi
}^{n}(x_{1},\cdots x_{n}),
\end{equation}%
where $\pi $ runs through the set $\Pi _{n}$ of all partitions of $n$
positive integers. Each partition $\pi $\ is characterized by $n$-positive
natural numbers $\{m_{1},m_{2},\cdots ,m_{n}\},$ which satisfy a Diofantine
equation%
\begin{equation*}
1\cdot m_{1}+2\cdot m_{2}+3\cdot m_{3}+\cdots n\cdot m_{n}=n,
\end{equation*}%
which in fact is a Frobenious equation. The symmetry condition is consistent
with $w_{n}(\pi )=w_{n}(\{m_{i}\}).$ Therefore, we can write%
\begin{equation}
P_{n}(x_{1},\cdots x_{n})=\sum_{\{m_{i}\}}w_{n}(\{m_{i}\})\mathcal{P}%
_{\{m_{i}\}}^{(n)}(x_{1},\cdots x_{n}).  \label{QueenConjunta}
\end{equation}%
By using the mathematical principle of induction, it is possible to obtain%
\begin{equation}
w_{n}(\{m_{i}\})=\frac{(n+\lambda )}{\lambda }\prod_{_{i=1}}^{n}\frac{%
[(i-1)!\lambda ]^{m_{i}}}{(i+\lambda )},
\end{equation}%
which can be rewritten as%
\begin{equation}
w_{n}(\{m_{i}\})=\frac{\Gamma (\lambda )}{\Gamma (n+\lambda )}%
\prod_{i=1}^{n}[\lambda (i-1)!]^{m_{i}}.
\end{equation}%
On the other hand, for each set of numbers $\{m_{i}\}$ the corresponding
probability reads%
\begin{eqnarray*}
\mathcal{P}_{\{mi\}}^{(n)}(x_{1},\cdots x_{n}) &=&\sum_{\{\chi \}}\Big{\{}%
\prod_{i_{1}=1}^{m_{1}}P_{1}(\chi _{i_{1}}) \\
&&\!\!\!\!\!\!\times \prod_{i_{2}=1}^{m_{2}}P_{1}(\chi _{i_{2}})\delta (\chi
_{i_{2}}-\chi _{i_{2}}^{(1)}) \\
&&\!\!\!\!\!\!\!\!\!\!\!\!\!\!\!\!\!\!\!\!\!\!\times
\prod_{i_{3}=1}^{m_{3}}P_{1}(\chi _{i_{3}})\delta (\chi _{i_{3}}-\chi
_{i_{3}}^{(1)})\delta (\chi _{i_{3}}-\chi _{i_{3}}^{(2)}) \\
&&\!\!\!\!\!\!\!\!\!\!\!\!\!\!\cdots \times
\prod_{i_{n}=1}^{m_{n}}P_{1}(\chi _{i_{n}})\prod_{j=1}^{n-1}\delta (\chi
_{i_{n}}-\chi _{i_{n}}^{(j)})\Big{\}}.
\end{eqnarray*}%
The set of variables $\{\chi _{i}\}$ assume values over the set $%
\{x_{i}\}_{i=1}^{n}.$ In each product there are $m_{j}$ independent
variables $\chi _{i_{j}},$ each one having associated other different $j$
variables $\chi _{i_{j}}^{(j)}$ that, due to the delta-Dirac contributions,
assume the same value than $\chi _{i_{j}}.$ Hence, in the previous
expression we have in total $m_{1}+2m_{2}+3m_{3}+\cdots nm_{n}=n$ different
variables $\chi .$ The sum runs overs all possible set of variables $\chi
\rightarrow x$ that lead to a different contribution. Consequently, the sum $%
\sum_{\{\chi \}}$ has a number $\mathcal{N}(\{m_{i}\})$ of different terms,
where%
\begin{equation}
\mathcal{N}(\{m_{i}\})=\frac{n!}{m_{1}!(1!)^{m_{1}}m_{2}!(2!)^{m_{2}}\cdots
m_{n}!(n!)^{m_{n}}}.
\end{equation}%
This number follows by taking into account that $\mathcal{P}%
_{\{mi\}}^{(n)}(x_{1},\cdots x_{n})$ does not depends on the order that the
variables appear.

From the previous results, we can study the statistics of the sum variable
Eq. (\ref{Wdefinition}). For simplifying the notation $G_{1}(k)\rightarrow
G(k).$ From the multiple Fourier transform Eqs. (\ref{FourierSuma}) and (\ref%
{GwFinal}), it follows%
\begin{equation}
G_{w}^{(2)}(2k)=\frac{\lambda G^{2}(k)+G(2k)}{(1+\lambda )}.
\end{equation}%
Similarly,%
\begin{equation*}
G_{w}^{(3)}(3k)=\frac{\lambda ^{2}G^{3}(k)+3\times \lambda
^{2}G(k)G(2k)+2G(3k)}{(1+\lambda )(2+\lambda )}.
\end{equation*}%
In general, from Eq. (\ref{QueenConjunta}) we obtain%
\begin{equation}
G_{w}^{(n)}(nk)=\sum_{\{m_{i}\}}w_{n}(\{m_{i}\})\mathcal{N}%
(\{m_{i}\})\prod_{i=1}^{n}[G(ik)]^{m_{i}}.
\end{equation}%
This expression explicitly reads%
\begin{equation}
G_{w}^{(n)}(nk)=\frac{n!\Gamma (\lambda )}{\Gamma (n+\lambda )}%
\sum_{\{m_{i}\}}\prod_{i=1}^{n}\frac{1}{m_{i}!}\Big{[}\lambda \frac{G(ik)}{i}%
\Big{]}^{m_{i}}.
\end{equation}%
Now, we note that this expression can be write in terms of a complete Bell
polynomial $B_{n}(x_{1},\cdots x_{n}),$\ which is defined as \cite{pitman}%
\begin{equation}
B_{n}(\{x_{i}\})=n!\sum_{\{m_{i}\}}\prod_{i=1}^{n}\frac{1}{m_{i}!}\Big{(}%
\frac{x_{i}}{i!}\Big{)}^{m_{i}}.
\end{equation}%
Therefore, under the association $x_{i}=(i-1)!\lambda G(ik),$ we can write%
\begin{equation}
G_{w}^{(n)}(k)=\frac{\Gamma (\lambda )}{\Gamma (n+\lambda )}%
B_{n}(\{(i-1)!\lambda G(ik/n)\}).
\end{equation}%
The complete Bell polynomial can be written as a determinant of a matrix, $%
B_{n}(\{x_{i}\})=B_{n}(x_{1},\cdots x_{n})=\det \mathcal{B}_{n}\mathcal{,}$%
\begin{equation*}
\mathcal{B}_{n}\!=\!\left( 
\begin{array}{ccccccc}
x_{1} & \binom{n-1}{1}x_{2} & \binom{n-1}{2}x_{3} & \binom{n-1}{3}x_{4} & 
\cdots & \cdots & x_{n} \\ 
&  &  &  &  &  &  \\ 
-1 & x_{1} & \binom{n-2}{1}x_{2} & \binom{n-2}{2}x_{3} & \cdots & \cdots & 
x_{n-1} \\ 
&  &  &  &  &  &  \\ 
0 & -1 & x_{1} & \binom{n-3}{1}x_{2} & \cdots & \cdots & x_{n-2} \\ 
&  &  &  &  &  &  \\ 
0 & 0 & -1 & x_{1} & \cdots & \cdots & x_{n-3} \\ 
&  &  &  &  &  &  \\ 
0 & 0 & 0 & -1 & \cdots & \cdots & x_{n-4} \\ 
\vdots & \vdots & \vdots & \vdots & \ddots & \ddots & \vdots \\ 
0 & 0 & 0 & 0 & \cdots & -1 & x_{1}%
\end{array}%
\right) \!,
\end{equation*}%
giving an exact and compact expression for $G_{w}^{(n)}(k).$ It can be
evaluated for arbitrary characteristic functions $G(k).$ This urn model also
lead to a wide family of probability densities that departs from a delta
Dirac distribution, Eq. (\ref{DeltaIndependientes}).

\end{document}